\documentclass[prb,amsmath,amssymb,showpacs]{revtex4-1}
\usepackage{graphicx,psfrag,color,hyperref}
\usepackage{amssymb,latexsym,mathrsfs,mathrsfs}
\usepackage{float}
\usepackage{array}
\usepackage{setspace}
\usepackage{caption}
\usepackage{wrapfig}
\usepackage{titlesec}	
\usepackage{lipsum} 
\usepackage{varioref}
\usepackage{cleveref}
\usepackage{textcomp}
\usepackage{epstopdf}
\usepackage{csquotes}
\usepackage{xcolor}
\usepackage{diagbox}
\usepackage{makecell}
\usepackage[bottom]{footmisc}
\usepackage{soul}
\usepackage{mwe}
\usepackage{adjustbox}
\usepackage{multirow}
\AppendGraphicsExtensions{.tif}
\begin{document}

\title{Setup of high resolution thermal expansion measurements in closed cycle cryostats using capacitive dilatometers}
	
\author{Neeraj K. Rajak$^1$}	
\author{Neha Kondedan$^1$}
\author{Husna Jan$^1$}
\author{Muhammed Dilshah U$^1$}
\author{Navya S. D.$^1$}
\author{Aswathy Kaipamangalath$^{2,3}$}
\author{Manoj Ramavarma$^{2,3}$}
\author{Chandrahas Bansal$^4$}
\author{Deepshikha Jaiswal-Nagar$^{1}$*}
\affiliation{$^1$School of Physics, IISER Thiruvananthapuram, Vithura, Kerala-695551, India}
\affiliation{$^2$Materials Science and Technology Division, CSIR-National Institute for Interdisciplinary Science and Technology, Thiruvananthapuram-695019, India}
\affiliation{$^3$Academy of Scientific and Innovative Research (AcSIR), CSIR-Human Resource Development Centre, (CSIR-HRDC) Campus, Postal Staff College Area, Sector 19, Kamla Nehru Nagar, Ghaziabad, Uttar-Pradesh-201002, India}
\affiliation{$^4$School of Physics, University of Hyderabad, Hyderabad, Telangana, India, 500046, India}
*\large \textbf{Email}:deepshikha@iisertvm.ac.in ~~~~~~~~~~~~~~~~~~~~~~~~~~~~~~~~~~~~~~~~~~~~~~~~~~~~~~~~~~~~~~\today
\begin{abstract}
We present high resolution thermal expansion measurement data obtained with high relative sensitivity of $\Delta$L$/$L = 10$^{-9}$ and accuracy of $\pm$2$\%$ using closed cycle refrigerators employing two different dilatometers. The data quality is in excellent agreement with those obtained using wet liquid helium based systems, demonstrating great technological possibilities for future thermal expansion measurements in view of the depleting resource of liquid helium. The cryogenic environment was achieved using two different cryostats that use pulse tube and Gifford-Mcmahon coolers as the cryocoolers. Both the dilatometers employ a spring movement for achieving the parallel movement of the capacitor plates. $Dilatometer \#1$ was built in-house based on a published design while $dilatometer \#2$ was obtained commercially. Cell calibration for $dilatometer \#1$ was done using copper and minimal deviation of the cell effect from the published values were found. Linear thermal expansion coefficient $\alpha$ obtained using both dilatometers was evaluated using two different techniques, namely, numerical differentiation and derivative of a polynomial fit. The resultant $\alpha$ obtained for metals silver and aluminium showed excellent match with published values obtained on systems using wet cryostats. Finite element method simulations were performed for understanding the spring movement in each dilatometer using which the effect of different forces$/$pressures on the displacement of the spring was studied. Finally, we report thermal expansion measurements done on single crystals of two high temperature superconductors YBa$_2$Cu$_{3-x}$Al$_x$O$_{6+\delta}$ and Bi$_2$Sr$_2$CaCu$_2$O$_{8+x}$ along the c-axis and found very good match with published data obtained using wet liquid helium based cryostats.
\end{abstract}

\maketitle

\section{Introduction}
Thermal expansion is an important thermodynamic technique to probe phase transitions in solids \cite{touloukian,barron}. Volume thermal expansion coefficient, $\beta$, describes the change in volume of a material in response to a change in temperature while the linear thermal expansion coefficient, $\alpha$, describes the corresponding change in the length of a given material. For an isotopic material, $\beta$ = 3$\alpha$. Gr{\"u}neisen parameter \cite{gruneisen,landau}, defined as $\Gamma$ = $\frac{\beta V}{\kappa_T C_V}$ where $\kappa_T$ is the isothermal compressibility, relates the volume thermal expansion coefficient $\beta$ to specific heat $C_V$. Zhu et al. defined a generalised Gr{\"u}neisen parameter $\Gamma$ = $\frac{\alpha}{c_p}$ = -$\frac{1}{V_m T}$$\frac{\partial S/\partial p}{\partial S/\partial T}$ for probing quantum phase transitions where the Gr{\"u}neisen parameter is expected to diverge \cite{zhu}. Such divergences in thermal expansion have, in fact, been observed in quantum critical materials like Sr$_3$Ru$_2$O$_7$ \cite{gegenwart}, (C$_5$H$_{12}$N)$_2$CuBr$_4$ \cite{lorenz}, CeCoIn$_5$ \cite{zaum} etc. Similarly, thermal expansion has been used as an important technique for studying correlations with onset of superconductivity to structural transitions in the crystalline lattice \cite{meingast,kund,meingastbscco,nagel,lortz,anshukova,mukherjee}. In this regard, thermal expansion measurements in high temperature superconductors (HTSC)'s are very valuable since the transition from the normal Mott insulating state to the  superconducting state is known to be accompanied by a corresponding structural transition from the tetragonal state of Mott insulator to orthorhombic state of superconductors \cite{orenstein,manjuYBCO}. Additionally, thermal expansion measurements in the HTSC YBa$_2$Cu$_3$O$_{6+x}$ (YBCO) have shown that superconductivity in YBCO is not only effected by the oxygen content of the superconductor but also by the degree of oxygen order in the CuO chain layer \cite{nagel}. Since HTSC's are proposed to be governed by an underlying quantum critical point \cite{orenstein,castro,keimer,sterpetti}, thermal expansion measurements-a generalised Gr{\"u}neisen parameter, offer a very important tool to probe quantum criticality in HTSC's.\\    
Of the many available techniques to measure thermal expansion of solids, capacitive dilatometry is known for its ability to resolve relative length changes of $\Delta$L$/$L $<$ 10$^{-9}$, such that a resolution of $\Delta$L = 10$^{-2}${\AA} can be achieved in samples of lengths in the mm range \cite{barron,barron_ASM,white,pott,steinitz,kund1,kroeger,neumeierPRB,schmiedeshoff,neumeier,manna,kuchler}. In comparison, a resolution of $\Delta$L = 10$^{0}${\AA} is reached using optical methods \cite{daou} and that of $\Delta$L = 10$^{-1}${\AA} is reached using the piezocantilever technique \cite{park}. In the capacitive dilatometry technique, the sample to be measured and two parallel metal plates are in a configuration such that a length change $\Delta$L of the sample results in a change in the gap between two metal plates placed parallel to each other. A high resolution capacitance bridge is, then, used to measure the capacitance between the plates before and after the length change, represented as $C_i$ (initial capacitance) and $C_f$ (final capacitance) respectively. The change in length $\Delta$L is equal to the negative of the change in gap between the capacitor plates, $\Delta D$ = $\epsilon_0$$\epsilon_r$ A(1/$C_i$ - 1/C$_f$), where $\epsilon_0$ and $\epsilon_r$ are absolute and relative permittivity of the medium in between the plates and A is the area of the capacitor plate. For a capacitive dilatometer to be useful for studying thermal expansion at low temperatures and high magnetic fields it must have certain features: (1) low thermal mass, (2) relative insensitivity to magnetic fields, and most importantly (3) good resolution. Low thermal mass ensures that the dilatometer can be efficiently cooled, requiring minimum amounts of helium in a cryogen based system or energy spent by a cryocooler in a dry system. This can be achieved by reducing the dimensions of the capacitor plates and the flanges that hold the capacitor plates, sample platform etc.\\
Superconducting samples or the systems exhibiting quantum phase transitions are frequently millimeter sized single crystals. Their interesting thermal expansion characteristics are observed at low temperatures where the thermal expansion is quite small. So, one requires dilatometers that have high resolution, typically in the sub-angstrom resolution range, to ensure that the small changes in thermal expansion can be picked up by the dilatometer efficiently. The first description of such a miniaturized dilatometer was given by White \cite{white}, which was later followed by Pott and Schefzyk \cite{pott}, Swenson \cite{swenson} and many others \cite{rotter,schmiedeshoff,steinitz,kund,kroeger,neumeierPRB,neumeier,manna,kuchler,kund1}. The designs can be differentiated based on three important characteristics: (1) material used in different components of a given dilatometer (2) sample mounting architecture, and (3) capacitor plate movement. A dilatometer that can be used to perform measurements in high magnetic fields requires its parts to be manufactured with materials that are insensitive to magnetic fields. For its relative insensitivity to magnetic field, ease of machinability and well known thermal expansion characteristics, Oxygen free high conductivity (OHFC) copper has been used in Schmidshoff's dilatomter design \cite{schmiedeshoff}, while K{\"u}chler \textit{et al.} \cite{kuchler} used an alloy of beryllium and copper to construct their dilatometer. On the other hand, Neumeier et al. \cite{neumeier} constructed their dilatomter using quartz to minimise the contribution due to background as much as possible. The initial dilatmeter designs were such that the samples were placed between the plates of the capacitor putting severe restrictions on the sample shape or geometry \cite{white,pott}. However, designs of Schmiedeshoff et al. \cite{schmiedeshoff} and K{\"u}chler et al. \cite{kuchler} have an open architecture, wherein, the sample is placed outside the capacitor plates resulting in choice of sample shape size and mountings. Finally, parallel plate movement of the capacitor plates have been achieved both by Neumeier et al. \cite{neumeier} and K{\"u}chler \textit{et al.} \cite{kuchler} by using leaf springs.\\
Thermal expansion measurements using capacitive dilatometry typically use wet cryostats where liquid helium is used as a cryogen. The reason for this is the sub-angstrom level resolution that one needs to achieve in order to investigate low temperature phase transitions in small sized single crystals. Closed cycle refrigerators (CCR's), on the other hand, frequently suffer from noise imparted to the experimental probe via the mechanical vibrations arising from the moving part of the CCR \cite{radebaughPT,radebaugh,waele}, thereby, seriously limiting the usage of CCR's for thermal expansion measurements. However, with the rapid improvement in the CCR technology in the last 20 years, lower noise vibration levels have been achieved in the CCR's  \cite{radebaugh,waele}, giving hope for their usage in noise-sensitive measurements like thermal expansion and reducing the dependence on the ever depleting and expensive cryogen of liquid helium. In this work, we present thermal expansion measurements on two different dilatometers, one based on Schmiedeshoff's design \cite{schmiedeshoff} that was fabricated, assembled and set-up in house, and the other based on K{\"u}chler's design \cite{kuchler}, obtained commercially. Both the dilatometers were implemented in the widely used physicsl property measurement system (PPMS) from Quantum Design Inc. using two different kinds of CCR's, namely, Gifford–McMahon (GM) CCR and Pulse-Tube (PT) CCR. Their performance was checked with standard metal samples of copper, aluminium and silver, with well-known thermal expansion characteristics. We find that thermal expansion measurement data obtained using either of the CCR's and either of the dilatometers give excellent quality data, comparable to those obtained using wet liquid helium cryostats. However, comparison between the cryostats and dilatometers reveal that PT cooled CCR and K{\"u}chler's dilatometer give better data. We employed two different methods of calculation of $\alpha$ from the relative length change, the first being a simple numerical dervative, and the second a polynomial fit. Both the methods yield different levels of noise in the calculated $\alpha$ and we find it necessary to consider both the methods of calculating $\alpha$ to better understand the CCR's contribution to the thermal expansion. For a dilatometer to be functional in different applied forces as well as varying spring thickness, it is necessary to ensure that the displacement of the spring on application of a force on the sample resulting in the movement of the capacitor plate, due to a change in the length of the sample, is monotonic. This was studied computationally by performing a finite element method (FEM) simulation.  Finally, we measure thermal expansion of two different superconducting crystals YBa$_2$Cu$_{3-x}$Al$_x$O$_{6+\delta}$ (Al-YBCO) and Bi$_2$Sr$_2$CaCu$_2$O$_{8+x}$ (BSCCO-2212) along the crystallographic c-axis and find consistent data with published reports.
  
\section{Experimental setup}
\begin{figure}[H]
	\begin{center}
		\includegraphics[width=1.0\textwidth]{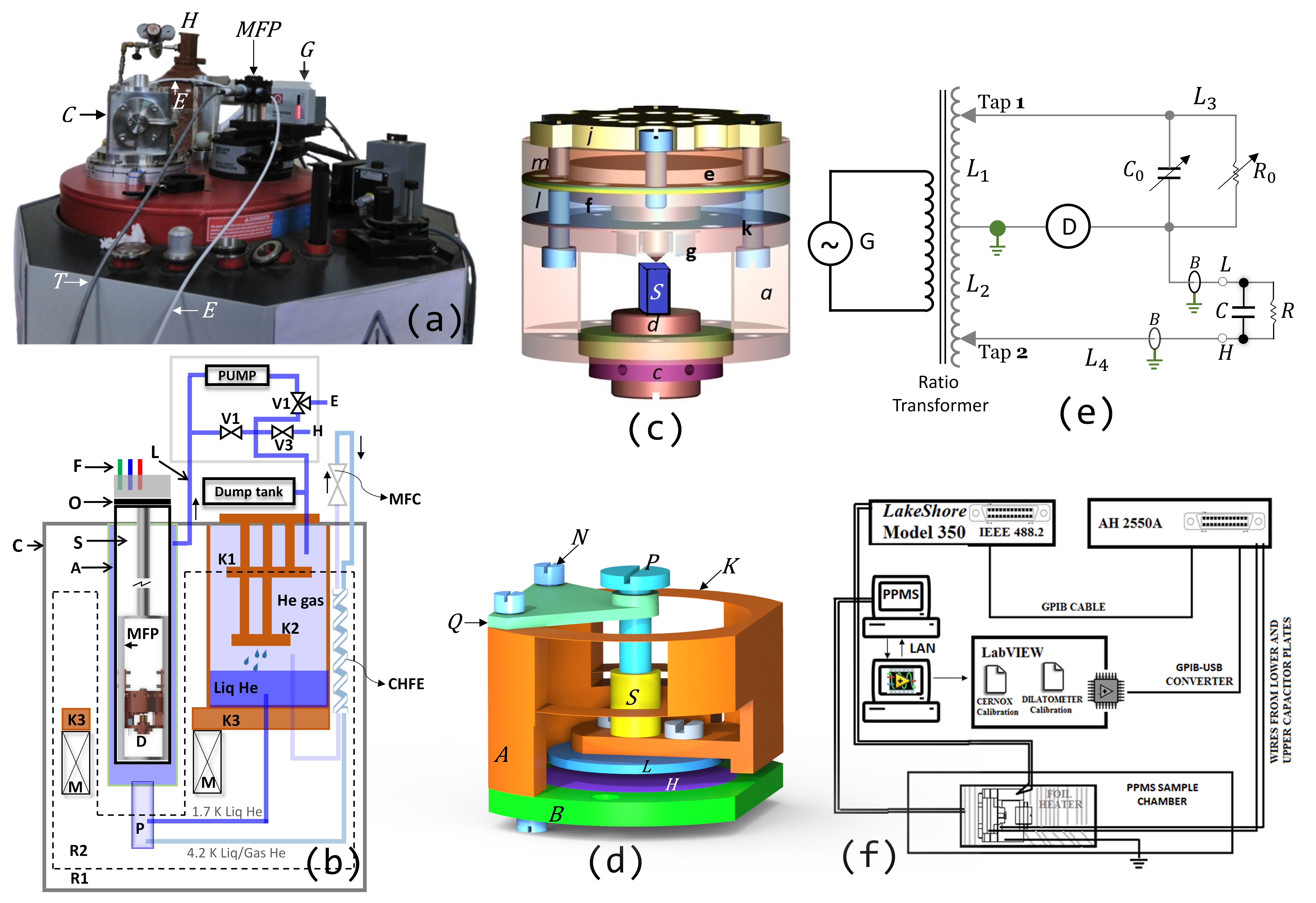}
		\caption{(colour online) (a) Thermal expansion set-up in a CCR based cryostat. H helium gas cylinder, MFP Top end of multi fuction probe, C cold head, E capacitance measuring cables and T temperature sensing cable. (b) Schematic of a PT cooled CCR. C outer wall of the cryostat, R1 outer vacuum jacket, R2 inner vacuum space, K1, K2 and K3 first, second and third stage of the cryocooler respectively, CHFE counter heat flow exchanger, MFC mass flow controller, M superconducting magnet, A annulus region of the main chamber, S sample space, MFP multi-function probe, F electrical feedthrough, O adapter, L annulus pumping line, V1, V2 and V3 bypass, exhaust and fill valves respectively, E exhaust line, H helium supply line and P pot. (c) and (d) are schematic of $dilatometer \#1$ and $dilatometer \#2$ respectively. (c) $a$ main flange, $c$ lock ring, $d$ sample platform, $S$ sample, $e$ upper capacitor plate, $g$ lower capacitor plate, $l$ shim, $m$ screws, $j$ mounting plate and $k$ spring (d) $S$ sample, $A$ main flange, $F$ lower flange, $L$ top capacitor plate, $H$ bottom capacitor plate, $P$ adjustment screw, $N$ nut and $Q$ adapter. (e) circuit used in the AH2500A capacitance bridge for the measurement of capacitance using the bridge balancing technique, C and R are unknown capacitance and resistance respectively, C$_0$ and R$_0$ fused silica capacitor capacitance and resistance respectively, G generator, D AC voltage detector, (f) Schematic of the electrical connections, instrument control and various components of the thermal expansion measurement system.}
		\label{fig:Schematics_Instrumentation}
	\end{center}
\end{figure}

The measurement setup used for thermal expansion measurements is shown in Fig. \ref{fig:Schematics_Instrumentation}. The setup consists of four different components: (1) dilatometer (2) cryostat (3) capacitance bridge and (4) temperature controller. We have used two different dilatometers in the present work. The first one, called as $dilatometer \#1$, was designed based on Schmiedeshoff's work \cite{schmiedeshoff}. Fig. \ref{fig:Schematics_Instrumentation} (c) and (d) show a computer realisation of $dilatometer \# 1$ and $dilatometer \# 2$ respectively made using CREO Parametric software. $Dilatometer \# 1$ consists of an upper capacitor plate $e$ that is fixed and a moving lower capacitor plate $g$. The two capacitor plates are separated by a shim $f$. As can be seen from Figs. \ref{fig:Schematics_Instrumentation} (c) and (d), both $dilatometer \#1$ as well as $dilatometer \#2$ have an open architecture design of sample mounting where the sample $S$ is placed outside the capacitor plates. Length changes brought in the sample due to a temperature variation is manifested as a change in the distance between the two capacitor plates brought about efficiently through a spring ($k$ in $dilatometer \#1$ and $K$ in $dilatometer \#2$). All the parts were fabricated using OFHC Copper and assembled together according to the details given in reference \cite{schmiedeshoff}. $Dilatometer \#2$, on the other hand, was obtained commercially from Kuechler Innovative Measurement Technology, Germany, which is based on the design of reference \cite{kuchler} and is shown in Fig. \ref{fig:Schematics_Instrumentation} (d). It has a fixed lower capacitor plate $H$ that is mounted onto the cell frame $A$, a movable upper capacitor plate $L$ and an adjustment screw $p$ using which the sample $S$ is fixed. The movement of the upper capacitor plate is brought about by a set of leaf springs $K$ machined out of BeCu. From the Fig. \ref{fig:Schematics_Instrumentation} (d), it is clear that the cell frame, the two capacitor plates as well as the spring are machined out of a single block of BeCu, resulting in a very high level of parallelism between the two capacitor plates.\\
The dilatometers were mounted on the multifunction probe (MFP) of Quantum Design Inc's dry PPMS which was used to provide the temperature platform. We used two different PPMS's that employ two different CCR's for providing the cryogenic environment: (1) PPMS-Evercool and (2)PPMS-Dynacool \cite{ppmsevercool,ppmsdynacool}. PPMS-Evercool uses GM CCR which is named as $CCR\#1$. On the other hand, PPMS-Dynacool employs a PT CCR which is named as $CCR\#2$ and is shown in Fig. \ref{fig:Schematics_Instrumentation} (a) where the magnet is not directly dipped in liquid helium. Either of the dilatometer ($dilatometer \#1$ or $dilatometer \#2$) was placed inside the cryostat using the MFP (see Fig. \ref{fig:Schematics_Instrumentation} (a and b)) that holds the dilatometer at its one-end. The MFP from Quantum Design is a versatile probe that comes with a connector at the bottom end which when plugged to the PPMS gives the temperature and the field, making it very user friendly probe \cite{ppmsevercool,ppmsdynacool}. It was, however, found that if the dilatometer is directly connected to the bottom connector of the MFP, the resulting capacitance data is extremely noisy with continuous spikes, arising possibly from the CCR.\\ 
To overcome this, we completely did away with the bottom connector of the MFP and provided our own temperature and capacitance connections at the top of the MFP as shown by wires T and E respectively in Fig. \ref{fig:Schematics_Instrumentation} (a). The electrical feedthroughs $F$ needed for making the connections are shown in Fig. \ref{fig:Schematics_Instrumentation} (b). For the $dilatometer \#1$, the electrical connections to the capacitor plates was done by first soldering two magnanin wires to the top (fixed) and bottom (movable) capacitor plates \cite{schmiedeshoff} and then soldering thin co-axial wires to the magnanin wires which were taken to the top of the MFP where it was soldered to the vacuum side of two hermetic co-axial Lemo connectors (see Fig. \ref{fig:Schematics_Instrumentation} (a and b)). $Dilatometer \#2$, on the other hand, already came with a set of thin co-axial wires that were directly connected to the Lemo connectors. The two capacitor plates and a ground wire were, then, connected to an Andeen-Hagerling capacitance bridge (Model AH2550A) for capacitance measurements in a three terminal capacitance fashion (Fig \ref{fig:Schematics_Instrumentation}(e)) to reduce stray capacitances of the cables, surroundings etc. \cite{pott}. In this scheme of measurement, a fixed frequency of 1 kHz from the generator G excites the ratio transformer comprising legs L$_1$ and L$_2$ each having taps to select precisely defined voltages that drive legs L$_3$ and L$_4$ of the bridge. L$_3$ comprises known variable capacitance C$_0$ and resistance R$_0$ generated from the bridge and is balanced to the unknown capacitance C and resistance R independently arising from the sample and comprise leg L$_4$ \cite{andeen-hagerling}.

Four wires carry the connections from a cernox temperature sensor that was attached to the dilatometer body, to a Lakeshore temperature controller (Model LS-350) (see Fig. \ref{fig:Schematics_Instrumentation} (f)). A LabVIEW program \cite{labview} was written to control the temperature, its scan rate, capacitance reading and display and finally data plotting continuously in both $CCR\#1$ and $CCR\#2$. Finally, in an effort to reduce possible noise vibrations from CCR's, we used a plastic adaptor and rubber O rings (shown as O in \ref{fig:Schematics_Instrumentation} (b)) to connect the MFP to the PPMS body.\\
All samples were cleaned with ethanol and iso-propanol and sanded lightly with sand paper of various grit sizes to make two parallel and smooth surfaces before mounting in a given dilatometer for measurement, except for Al-YBCO crystal which naturally has a cube shape. Hence, for this crystal no sanding was done and it was loaded after cleaning with ethanol. The temperature was sweeped in a continuous mode at a rate of 0.2 K/min. The measurement time for each capacitance data point was 8 seconds in averaging mode \cite{andeen-hagerling}. We used a combination of two software packages, (1) OriginLAB and (2) Octave/MATLAB \cite{octave} for data analysis, curve fitting and data plotting.
\begin{figure}
	\begin{center}
		\includegraphics[width=0.5\textwidth]{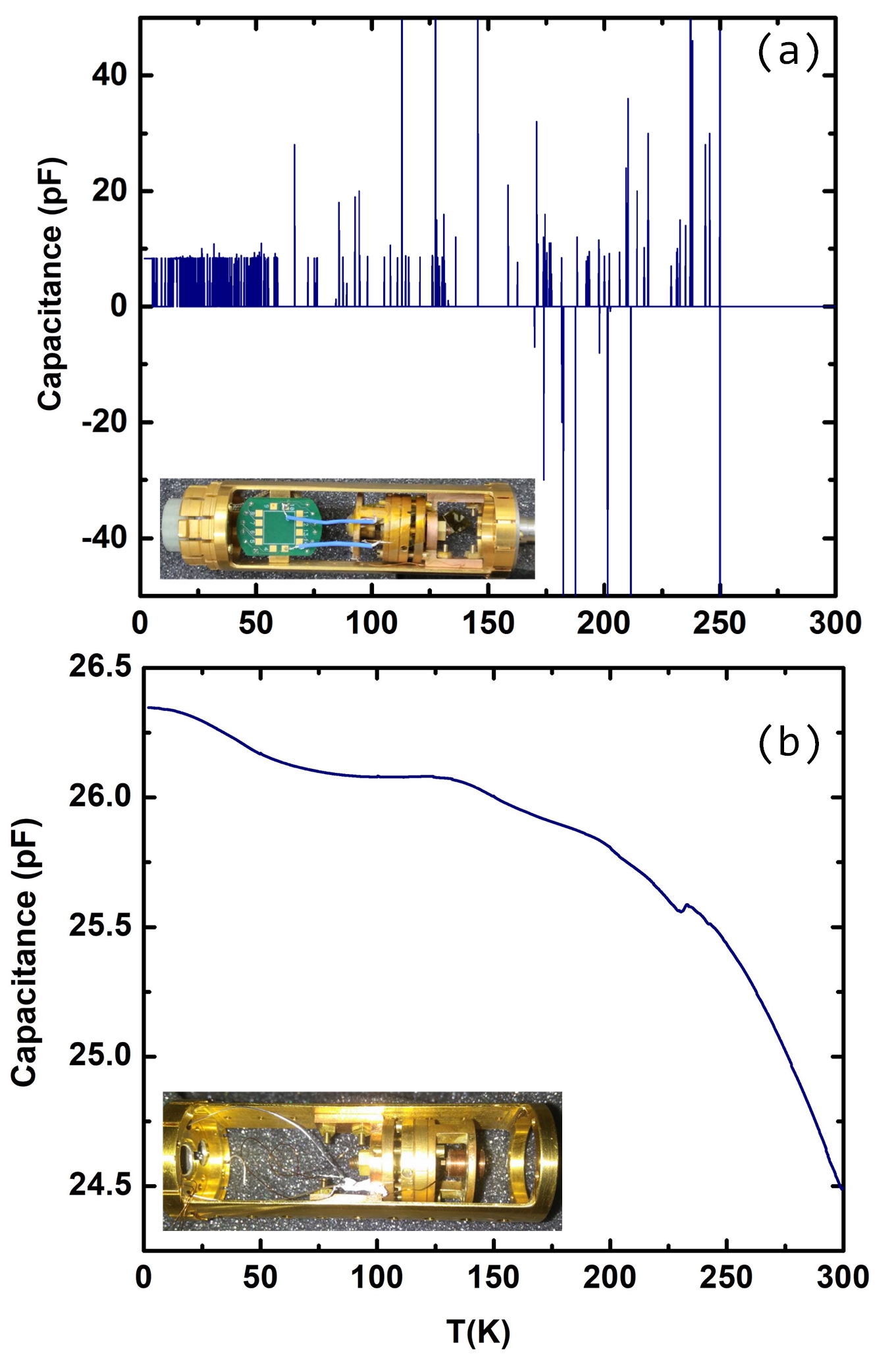}
		\caption{(colour online) (a) Plot of capacitance as a function of temperature T when the connections were taken from the bottom connector of the MFP. Inset shows the actual picture of the connection. (b) Same plot as in (a) when connections are taken from the top of MFP. Inset shows the actual picture of the $dilatometer \#1$ attached to MFP.}
		\label{fig:bottom_top}
	\end{center}
\end{figure}

\section{Measurement and analysis}
As described in the previous section, we took all the connections from the top of the MFP in an effort to improve the extremely bad quality data that was obtained on using the bottom connector of the PPMS. Main panel of Fig. \ref{fig:bottom_top} (a) shows a representative data that was taken on $dilatometer \#1$ using $CCR \#1$ on a copper sample. The data represents temperature variation of capacitance that is characterised by continuous and large spikes in capacitance as the temperature was varied. The inset of Fig. \ref{fig:bottom_top} (a) shows the bottom end of the MFP on which we connected the dilatometer using the connections provided by bottom connector of the MFP. However, when the connections were taken from the top of the MFP  as shown in Fig. \ref{fig:Schematics_Instrumentation} (a) and the inset of Fig. \ref{fig:bottom_top} (b), a fantastic improvement in the data quality was observed. Main panel of Fig. \ref{fig:bottom_top} (b) shows a representative capacitance vs. temperature data on the same $CCR \#1$ cryostat measured using the $dilatometer \#1$ on the copper sample. The data is marked by a continuous and smooth variation of capacitance with temperature with no intermittent jumps as was observed in Fig. \ref{fig:Schematics_Instrumentation} (a).
\subsection{Calibration and test at ambient temperature and pressure}	
Before mounting the samples on the dilatometers and loading them inside the CCR's for temperature dependent capacitance measurements, it was mandatory to do some room temperature tests to check for parallelism of the capacitor plates. The room temperature test involved measuring the capacitance, $C$, as a function of distance between the plates, $D$. For an ideal parallel placed capacitor with a surface area $A$, the capacitance is given by:
\begin{center}
	\begin{equation}
	\label{eq:ideal capacitor}
	C = \frac{\epsilon_0 \epsilon_r A}{D}
	\end{equation}
\end{center}

where $\epsilon_0$ = 8.8542 $\times$ 10$^{-12}$ $F/m$ is the permittivity of vacuum, $\epsilon_r$ is the dielectric constant of the medium between the parallel capacitor plates. If the dilatometers do have the ideal parallel placed capacitor geometry, the experimentally obtained $C$ vs. $D$ curve should follow equation \ref{eq:ideal capacitor}. 

\subsubsection{Dilatometer $\#$1}
In order to test the validity of the equation \ref{eq:ideal capacitor} and check for possible deviations from the ideal geometry, we performed a $C$ vs. $D$ test at room temperature and ambient pressures for $dilatometer\#1$. In this dilatometer, the change in distance between the plates is achieved by first shorting the sample platform $d$ to the lower capacitor plate $g$ and then rotating the sample platform by an angle $\theta$ to achieve a linear motion of the sample platform which eventually pushes the lower capacitor plate closer to or farther from the fixed upper capacitor plate $e$ depending on the direction of rotation. The precise amount by which the distance between the capacitor plates changes was measured using the formula in equation \ref{eq:platform-rotation}, wherein the symbol $\theta$ represents the angle by which the sample platform is rotated, $\theta_{max}$ is the value of the angle $\theta$ where the two capacitor plates short and $w$ is a constant which represents the pitch of the sample platform thread having a value of 882 nm per degree of rotation.

\begin{figure}
	\begin{center}
		\includegraphics[width=0.7\textwidth]{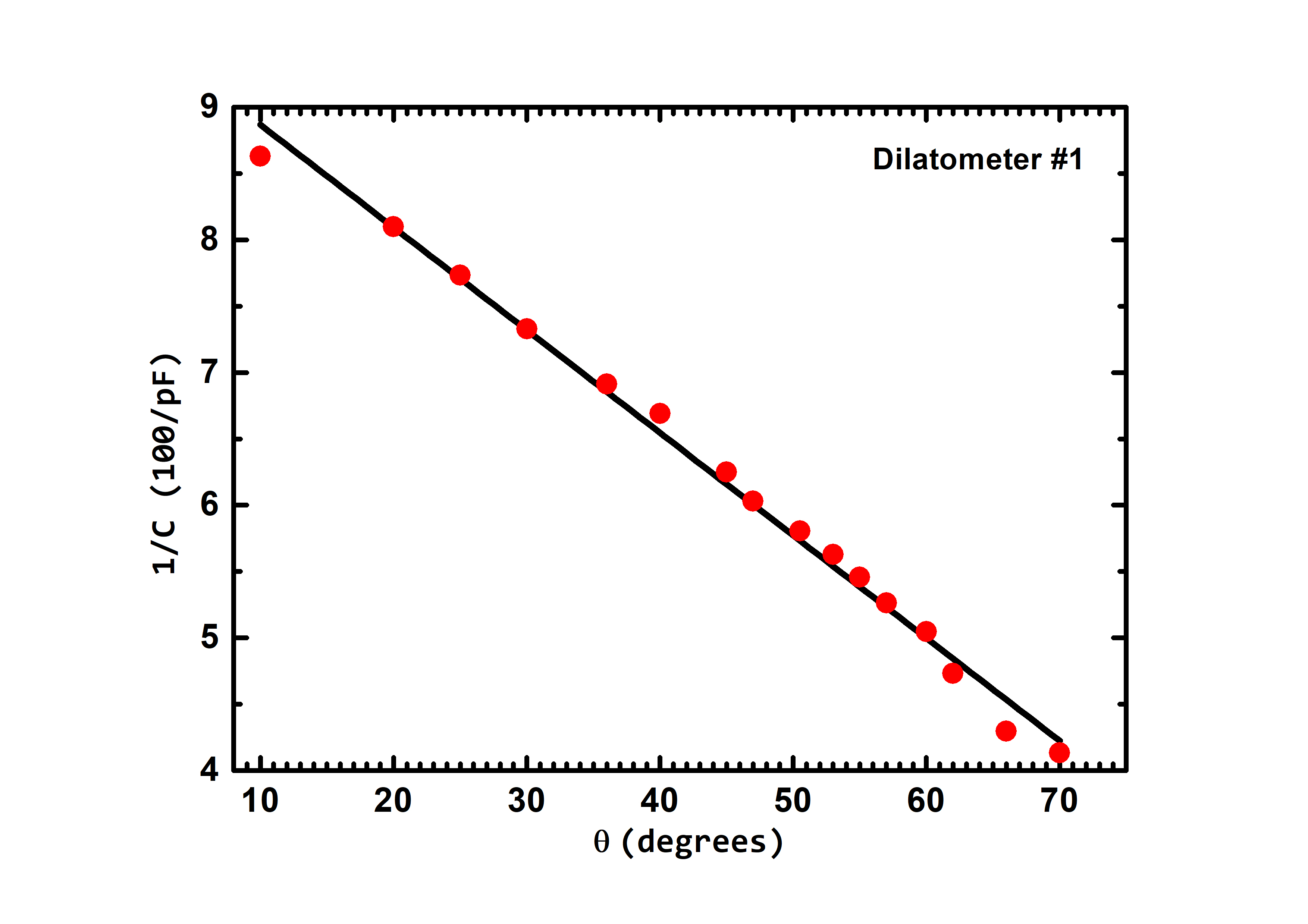}
		\caption{(colour online) Variation of inverse of capacitance (1$/$C) with the angular movement of the sample platform measured in angle $\theta$ at room temperature performed for the $dilatometer\#1$. Filled red circles correspond to data points while the black line is a straight line fit to equation \ref{eq:C-theta} (see text for details).}
		\label{fig:Room_Temperature_Test}
	\end{center}
\end{figure}
 
 \begin{center}
	\begin{equation}
	\label{eq:platform-rotation}
	D = (\theta_{max} - \theta) \times w 
	\end{equation}
\end{center}
 Substituting equation \ref{eq:platform-rotation} in equation \ref{eq:ideal capacitor} and assuming the dielectric constant of air $\epsilon_r$ as 1, we get:
\begin{center}
	\begin{equation}
	\label{eq:C-theta}
	\frac{1}{C} = - \bigg(\frac{w}{\epsilon_0 A}\bigg)\theta + \bigg(\frac{\theta_{max}\times w}{\epsilon_0 A}\bigg)
	\end{equation}
\end{center}
The equation \ref{eq:C-theta} gives $w / \epsilon_0A$ as the slope of the plot of $1/C$ vs. $\theta$ curve, which must be a straight line. We fitted this equation in a least square manner to a measured variation of $1/C$ vs. the variable $\theta$ shown in Fig. \ref{fig:Room_Temperature_Test}. It can be seen from Fig. \ref{fig:Room_Temperature_Test} that a straight line fits the data (shown as red filled circles) very well in the 12 pF$\leq C \leq$ 21 pF range. Deviation from linearity in the higher angle range may be arising due to non-perfect parallel arrangement of the capacitor plates \cite{schmiedeshoff}. The slope obtained from the straight line fit in the 12 pF$\leq C \leq$ 21 pF range is 0.07741 pF$^{-1} \theta ^{-1}$. So, the area of the capacitor plate obtained from the fit is 1.3338 $\times$ 10$^{-4}$m$^2$ which is $\sim$ 0.05$\%$ of the actual area of the capacitor plate (1.2711 $\times$ 10$^{-4}$ m$^2$ (d = 12.725 mm)). This difference in the actual value and the obtained value is consistent with effects like edge effects for circular capacitor plates separated by a small gap (geometry of $dilatometer \#1$) \cite{heerens}, roughness and curvature of capacitor plates etc. \cite{pott}. 

\subsubsection{Dilatometer $\#$2}
For this dilatometer, the change in the length of the sample, $\Delta$L, is achieved by measuring the change in the capacitance $C$ from the starting capacitance $C_0$ as follows:
\begin{center}
	\begin{equation}
	\label{eq:kuchler}
	\Delta L = \epsilon_0 \pi r^2\frac{C-C_0}{C\cdot C_0}\bigg(1-\frac{C \cdot  C_0}{C_{max}^2}\bigg)
	\end{equation}
\end{center}
where $r$ is the radius of the capacitor plate (= 7 mm) and $C_{max}$ is the value of the capacitance at which the two capacitor plates short. Equation \ref{eq:kuchler} was derived by Pott and Schefzyk \cite{pott} by incorporating a slight tilt between the capacitor plates. The change in the length, $\Delta$L, was obtained by using a dial gauge of resolution 1$/$100 mm while the capacitance was measured using a commercial capacitance bridge. By tightening the adjustment screw $p$, a graph of $\Delta$L vs. $C$ was obtained for different values of shorting capacitance $C_{max}$. This calibration chart was provided to us by the manufacturer. We have operated our $dilatometer \#2$ between the range of values of $C_0$ varying between 12 pF to 14 pF which corresponds to a length change of 8.7372 $\mu$m for a capacitance change of 1 pF for a sample of length 1mm. In this range of capacitance values, a parallel plate geometry assumption holds true \cite{kuchler}. $Dilatometer \#2$ has incorporated a parallelogram suspension mechanism \cite{steinitz,kuchler} for achieving the parallelism between the capacitor plates. We have studied this as well as other aspects of such a design by performing a FEM simulation, the details of which are presented in section 3.

\subsection{Empty cell effect}
\begin{figure}[H]
	\begin{center}
		\includegraphics[width=0.45\textwidth]{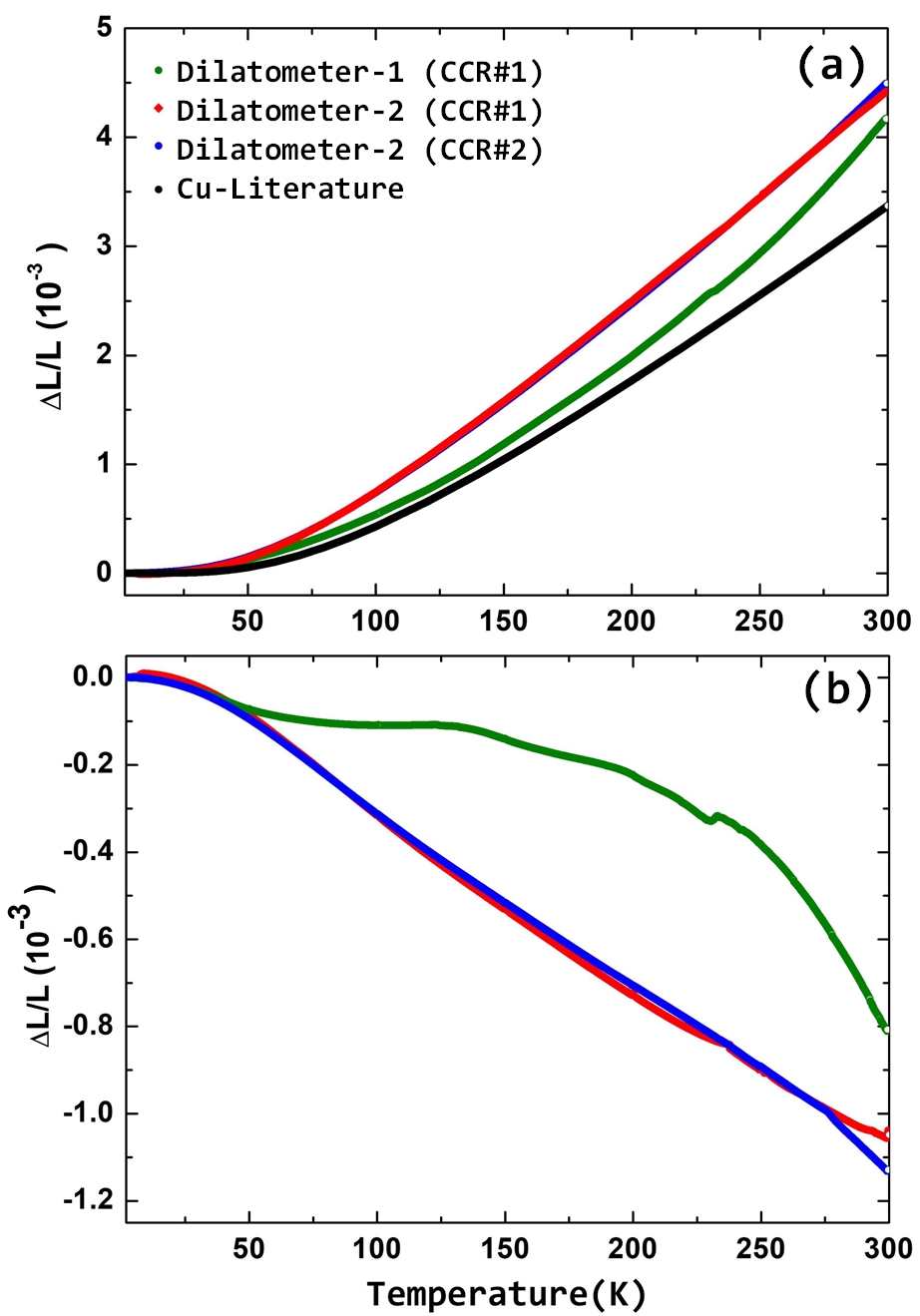}
		\caption{(colour online) (a) Plot of $\big(\Delta L/L_0 \big)^{cell}$ obtained for $dilatometer \#1$ on $CCR\#1$ (green),  $dilatometer \#2$ on $CCR\#1$ (red) and $dilatometer \#2$ on $CCR\#2$ (blue). Black curve is $\big(\Delta L/L \big)^{Cu}_{literature}$ taken from reference \cite{kroeger}. (b) Plot of the empty cell effect $\big(\Delta L/L_0 \big)^{emptycell}$ obtained on a copper sample for the three configurations as in (a).}
		\label{fig:Cell_Effect}
	\end{center}
\end{figure}

In a thermal expansion or magnetostriction measurement, the sample as well as the various components of the dilatometer cell undergo changes in their dimensions in response to a change in temperature or magnetic field. This means that the measured length change of a sample, $\Delta L^{sample}_{meas}$, is the difference between the change in length of the sample, $\Delta L^{sample}$ and a background contribution arising from the change in length of the dilatometer cell, $\Delta L^{cell}$, known as the cell effect \cite{schmiedeshoff,kuchler}. The cell effect can be determined by performing a thermal expansion measurement on a reference sample, whose thermal expansion characteristics are well known. For both $dilatometer \#1$ as well as $dilatometer \#2$, the sample platform itself could be used as reference samples. This was possible since $dilatometer \#1$ was fabricated from OFHC copper whose thermal expansion is well known and for $dilatometer \#2$ which was machined from Cu$_{1-x}$Be$_x$ with a low beryllium content of 1.84, copper's thermal exapansion data could be used \cite{kuchler}. For the measurements reported in this work, we have used a copper polycrystal which was cut in the shape of a cylinder of diameter 2 mm and height 2.8 mm. The expansion measurement performed with the reference copper sample yields the empty cell effect, $\Delta L^{empty cell}$, which is the difference of the cell length change $\Delta L^{cell}$ from the reported literature values of length change of copper, $\Delta^{Cu}_{literature}$:
\begin{center}
	\begin{equation}
	\label{eq:cell-effect1}
	\Delta L_{meas}^{Cu} = \Delta L^{empty cell} = \Delta L^{Cu}_{literature} -\Delta L^{cell}
	\end{equation}
\end{center}
The above equation comes from the fact that the only part of the cell frame which contributes to the measured length change is a piece of the frame which has exactly the same length as the sample being measured. Therefore, the relative length change of any sample, $\Delta L^{sample}$, is the sum of the measured length change of the sample, $\Delta L^{sample}_{meas}$, with the length change corresponding to the calibrated cell:
\begin{center}
	\begin{equation}
	\begin{split}
	\label{eq:cell-effect2}
	\Delta L^{sample}  = \Delta L^{sample}_{meas} + \Delta L^{cell},~~~~~~~~~~~~~~~~~~~~~~\\
	~~~~~~~ = \Delta L^{sample}_{meas} - \Delta L^{empty cell} + \Delta L^{Cu}_{literature}
	\end{split}
	\end{equation}
\end{center}  
The relative length change of the sample normalised to its initial length L$_0$ is, then, given by
\begin{center}
	\begin{equation}
	\label{eq:cell-effect3}
	\bigg(\frac{\Delta L}{L_0}\bigg)^{sample} = \frac{\Delta L^{sample}_{meas} - \Delta L^{empty cell}}{L_0} + \bigg(\frac{\Delta L}{L}\bigg)^{Cu}_{literature}
	\end{equation}
\end{center}
Figure \ref{fig:Cell_Effect} (a) shows the length change of the cell $\big(\Delta L/L_0 \big)^{cell}$ for $dilatometer \#1$ on $CCR \#1$ (green curve), $dilatometer \#2$ on $CCR \#1$ (red curve) and $dilatometer \#2$ on $CCR \#2$ (blue curve). Unfortunately, $CCR \#1$ broke down before we could measure $dilatometer \#2$ on it, so we do not have this data with us presently and is the subject matter for future work. The literature values of the relative length change $\big(\Delta L/L \big)^{Cu}_{literature}$ and the corresponding coefficient of thermal expansion, $\alpha$, for a pure copper sample was taken from the reference \cite{kroeger}. 
Fig. \ref{fig:Cell_Effect} (b) shows the calculated values of the relative length change corresponding to empty cell effect, $\big(\Delta L/L_0 \big)^{emptycell}$, using the equation \ref{eq:cell-effect1}. The colour code represents the same configuration as described in Fig. \ref{fig:Cell_Effect} (a). The first observation to be made from the data plotted in Fig. \ref{fig:Cell_Effect} (a) is that the data quality is very good, similar to the copper reference data (black curve) measured on a wet cryostat using liquid helium as a cryogen \cite{kroeger}. This observation, then, suggests that we have been able to mitigate the noise problems arising in dry cryogen free cryostats to such a level that the resultant data quality is very similar to that obtained from wet cryostats. This is very heartening in the light of the fact that liquid helium is a rapidly depleting cryogen, and, hence, one can reduce one's dependence on it for doing extremely sensitive experiments like thermal expansion requiring sensitivity of the order of  $\Delta$L$/$L $<$ 10$^{-9}$. From Fig. \ref{fig:Cell_Effect} (a), it can also be observed that $\big(\Delta L/L_0 \big)^{cell}$ is the same for both the dilatometers in both the cryostats till $\sim$ 90 K above which they differ in magnitudes. However, the temperature variation is smooth and monotonic for both. The data for $dilatometer \#2$ was found to be independent of the cryostat used since the two curves corresponding to $CCR \#1$ (red) and $CCR \#2$ (blue) were found to overlap.\\ 
K{\"u}chler et al. \cite{kuchler} found a temperature independent empty cell effect $\big(\Delta L/L_0 \big)^{emptycell}$ till $\sim$ 200 K on a scale of 0.5 $\cdot$ 10$^{-3}$ when measured in a wet PPMS/exchange gas cryostat. If we compare our $\big(\Delta L/L_0 \big)^{emptycell}$ data shown in Fig. \ref{fig:Cell_Effect} (b), on K{\"u}chler et al.'s scale \cite{kuchler}, the empty cell effect is nearly temperature independent till $\sim$ 280 K for $dilatometer \#1$ measured in $CCR \#1$. In comparison, for the $dilatometer \#2$ the temperature independence is over a smaller temperature interval of $\sim$ 150 K, comparable to what was obtained by K{\"u}chler et al. \cite{kuchler} in their wet cryostat.

\subsection{Thermal expansion of Aluminium}
To check the functioning of the dilatometers, it is necessary to first calibrate it with other metals having well known thermal expansion charateristics. To do this, we performed thermal expansion measurements on an aluminium polycrystal which was cut into the shape of a cylinder of diameter 2 mm and height 3 mm for $dilatometer \#1$ and diameter 2 mm and height 2.57 mm for $dilatometer \#2$.
\begin{figure}[H]
	\begin{center}
		\includegraphics[width=0.97\textwidth]{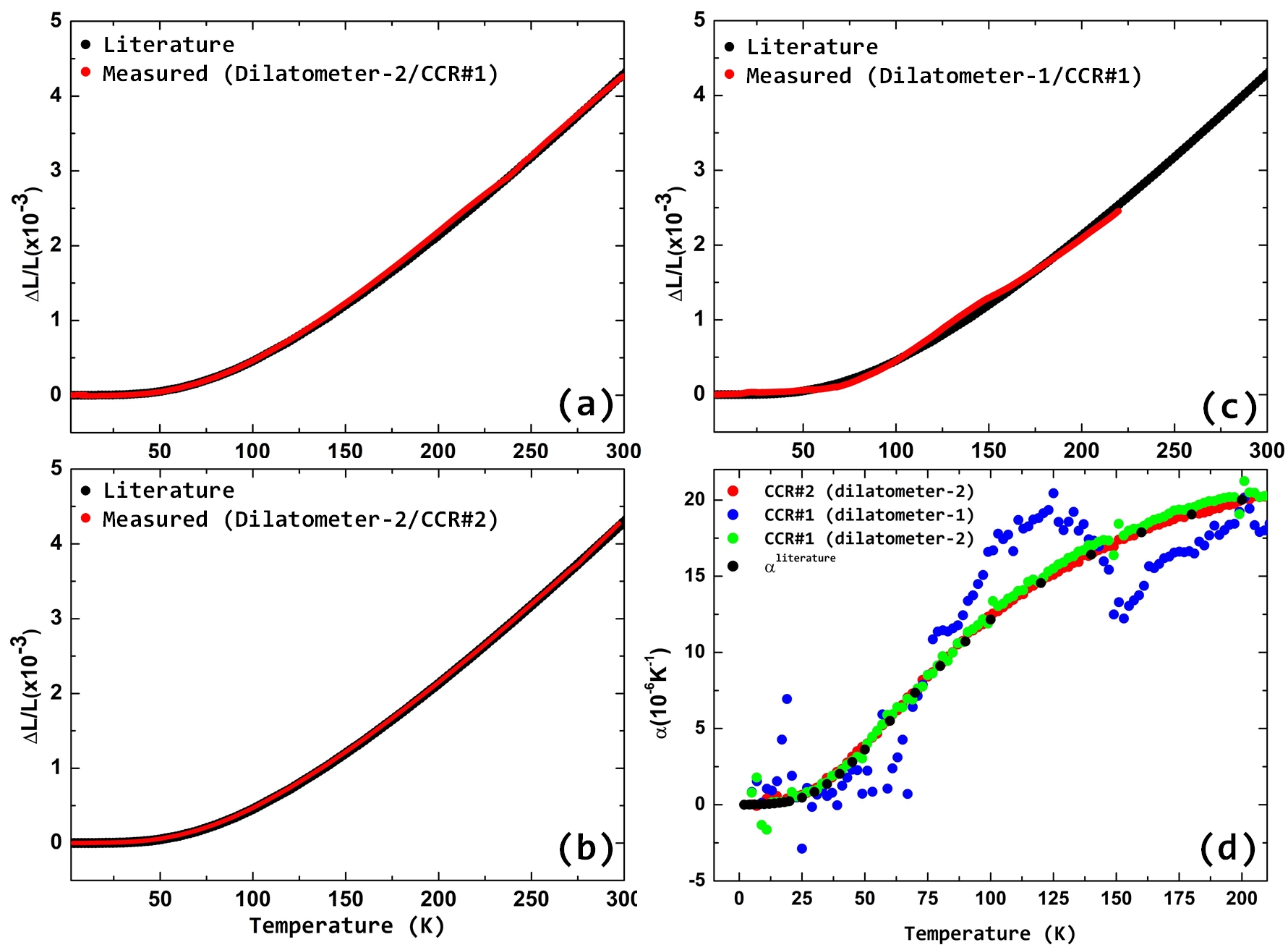}
		\caption{(Colour online) Filled red circles represent temperature variation of $\Delta$L$/$L values of an aluminium polycrystal obtained from three different sets of measurements performed with $dilatometer \#2$ on (a) $CCR\#1$ and (b) $CCR\#2$ while (c) represents $\Delta$ L$/$L values obtained with $dilatometer \#1$ on $CCR\#1$. Black filled circles in each panel represent literature values for the relative length change as a function of temperature extracted from \cite{kroeger}. (d) represents the calculated linear thermal expansion coefficient $\alpha$ for $dilatometer \#2$, $CCR\#1$ (green curve); $dilatometer \#2$, $CCR\#2$ (red curve) and $dilatometer \#1$, $CCR\#1$ (blue curve). Black filled circles represent the extracted literature data from \cite{kroeger}.}
		\label{fig:aluminium}
	\end{center}
\end{figure}
The values of the relative change in length $\Delta$L$/$L measured with $dilatometer \#2$ using $CCR \#1$ and $CCR \#2$ are shown by red curves in Fig. \ref{fig:aluminium} (a) and (b) respectively, while Fig. \ref{fig:aluminium} (c) shows the same data with $dilatometer \#1$ using $CCR \#1$. Black continuous curve in each figure is the literature data taken from reference \cite{kroeger}.
It can be seen that while the $\Delta$L$/$L curve obtained with $dilatometer \#1$ using $CCR \#1$ (see Fig. \ref{fig:aluminium} (c)) shows a slight deviation from the literature curve $\sim$ 140 K, the curves obtained with $dilatometer \#2$ using either of the CCR's (see Figs. \ref{fig:aluminium} (a) and (b)) show an excellent match with the reference curve.\\ 
Coefficient of thermal expansion $\alpha$ obtained for aluminium for the three different setups, as described above, are shown in Fig. \ref{fig:aluminium} (d) (green, red and blue), while the literature values of $\alpha$ oatained from reference \cite{kroeger} is shown in the black curve of Fig. \ref{fig:aluminium} (d). As is evident from the graph, the calculated values of $\alpha$ for $dilatometer \#2$ on either of the CCR's match with the literature values quite well. For the $dilatometer \#1$, on the other hand, the matching is not that good with large deviations from the literature values. 

\subsection{Calculation of $\alpha$}
The coefficient of thermal expansion $\alpha$ for a sample with an initial length L$_0$ is defined as:
\begin{equation}
	\label{eq:alpha}
	\alpha  = \frac{1}{L_0}\frac{dL}{dT} 
\end{equation}
Using equation \ref{eq:cell-effect3}, we get
\begin{equation}
	\label{eq:alphasample}
	\alpha_{sample}  = \frac{1}{L_0}\frac{dL}{dT} \bigg|^{sample}_{meas} - \frac{1}{L_0}\frac{dL}{dT} \bigg|^{emptycell} + \alpha_{literature}^{Cu} 
\end{equation}
where dL is the change in length in response to a change in temperature dT. For a given sample, the experimental data consists of discreet data points, so the derivatives required in equation \ref{eq:alphasample} can be calculated using either numerical derivative or derivative of a polynomial fitted to the data.

\subsubsection{Numerical derivative}
\begin{figure}[H]
	\begin{center}
		\includegraphics[width=0.45\textwidth]{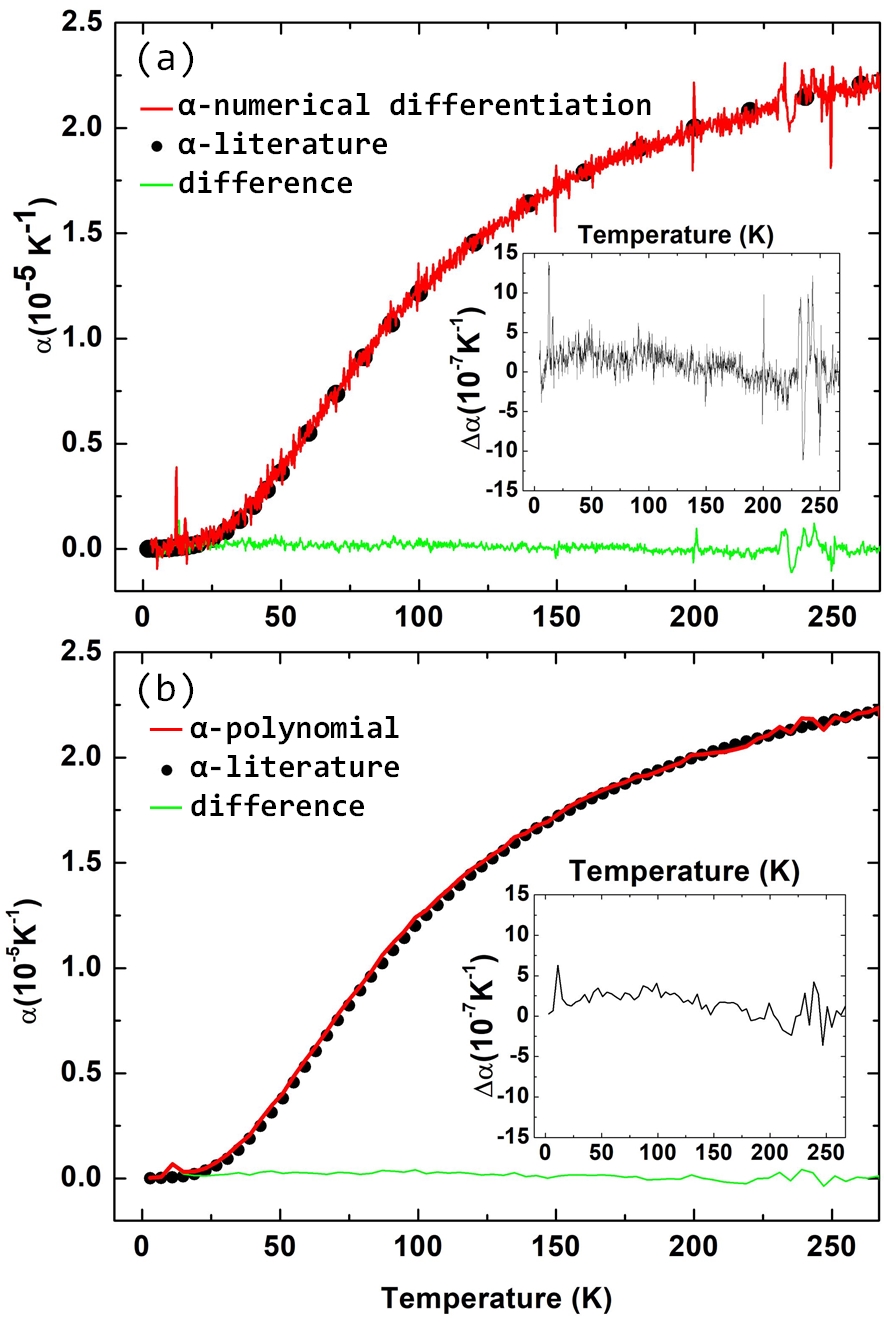}
		\caption{(colour online) Red continuous curves show linear thermal expansion coefficient $\alpha$ calculated for aluminium polycrystal using two different techniques (a) numerical differentiation of the data in Fig. \ref{fig:aluminium} and using the equation \ref{eq:numerical-derivative}. (b) Polynomial fit to the same data, where a polynomial of order 3 is fitted to every 10 data points and a derivative of the resultant polynomial is taken to calculate the value of $\alpha$. Black filled circles in each curve indicate the literature values of $\alpha$ obtained from Kroeger et al. \cite{kroeger}.}
		\label{fig:alpha-numerical-polynomial}
	\end{center}
\end{figure}

Numerical derivatives at various values of temperatures $T_i$ is defined with the help of equation \ref{eq:numerical-derivative}. 
	\begin{equation}
	\label{eq:numerical-derivative}
	\left[\frac{dL}{dT}\right]_{T = T_i} = \frac{1}{2}\left[\frac{L_{i+1} - L_i}{T_{i+1} - T_i} + \frac{L_i - L_{i-1}}{T_i - T_{i-1}} \right]
	\end{equation}
where $L_{i-1}$, $L_{i}$ and $L_{i+1}$ are the values of the length at the i-1$^{th}$, i$^{th}$ and i+1$^{th}$ temperature values respectively. The numerical differentiation was done using the Matlab console of the Origin package \cite{origin}. Fig. \ref{fig:alpha-numerical-polynomial} (a) shows the temperature variation of 
$\alpha$ calculated in this manner for an aluminium sample, whose $\Delta L/L$ was shown in Fig. \ref{fig:aluminium} (b) above. Literature values obtained from \cite{kroeger} are shown by black filled circles in Figs. \ref{fig:alpha-numerical-polynomial} (a) and (b). It can be seen that the calculated values of $\alpha$ is noisy but match the literature values over the entire temperature range. The data is characterised by few glitches with delta-function-like features in the high temperature range and one in the low temperaures $\sim$ 10 K. Green curve in Fig. \ref{fig:alpha-numerical-polynomial} (a) shows a difference curve, $\Delta \alpha$ which was obtained by subtracting the obtained values of $\alpha$ from the literature values.  

The average value of the absolute difference $|\Delta \alpha|$ was found to be 2 $\cdot$ 10$^{-7}$, so the relative error from the literature value is $\sim$ 1$\%$ shown more clearly on an expanded scale in the inset of Fig. \ref{fig:alpha-numerical-polynomial} (a). These values are consistent with typical values reported in \cite{schmiedeshoff,kuchler}.

\subsubsection{Polynomial fit}
In an effort to remove the glitches that appear in $\alpha$ data due to the numerical differentiation method described above, we performed a polynomial fit on the same data. To do this, a smooth polynomial fit was done on the quantities $\Delta L^{sample}_{meas}$ and $\Delta L^{emptycell}$ prior to calculation of derivatives. The goal was to find coefficients $\beta_i$'s corresponding to a polynomial, P of the type P = $\beta_0 + \beta_1 T + \beta_2 T^2 + \beta_3 T^3 + ... + \beta_n T^n$. This can be achieved via a least square fit to the dataset either in the entire temperature range, or by dividing the entire data range in sets of $m$ data points, so as to obtain $N/m$ different polynomials for $N$ total number of observations. We found that although it was possible to fit the data in the entire temperature range with a single polynomial, the fitted polynomial was not able to capture the features in data corresponding to small temperature changes. For this reason, the fits were performed using the latter approach, wherein, we took a set of 6 to 10 data points ($m \in [6,10]$) and fitted the $N/m$ set of points to a polynomial of degree $\le$ 3. A polynomial fit performed in such a fashion on the aluminium sample is shown in figure\ref{fig:alpha-numerical-polynomial}(b). It can be immediately observed that the glitches in the data which were present in Fig. \ref{fig:alpha-numerical-polynomial} (a) are considerably suppressed and the data presents a smooth variation of $\alpha$, similar to what has been reported in the literature \cite{kroeger}. The relative error, as shown by the green curve in Fig. \ref{fig:alpha-numerical-polynomial} (b) has also reduced with a reduced absolute average value $|\Delta \alpha|$ of 2 $\times$ 10$^{-7}$ as compared to that for the numerical derivative method. So, we have calculated the linear coefficient of thermal expansion $\alpha$ of aluminium icorporating the three different set-up's described above in Fig. \ref{fig:aluminium} (d), using this technique of polynomial fit. 

\subsection{Calculation of $\Delta L/L$  and $\alpha$ for silver}
From the results obtained for thermal expansion measurements performed on aluminium on the two dilatometers $dilatometer \#1$ and $dilatometer \#2$, and two different closed cycle refrigerators $CCR\#1$ and $CCR\#2$ (see Fig. \ref{fig:aluminium}) (d), we can clearly conclude that the dilatometer of choice for a thermal expansion measurement on a closed cycle cryostat is $dilatometer \#2$ (K{\"u}chler's dilatometer) and either of the closed cycle refrigerators, $CCR\#1$-Gifford–McMahon based PPMS-Evercool or $CCR\#2$-Pulse tube based PPMS Dynacool. As an additional test of the good performance of 
\begin{figure}[H]
	\begin{center}
		\includegraphics[width=0.45\textwidth]{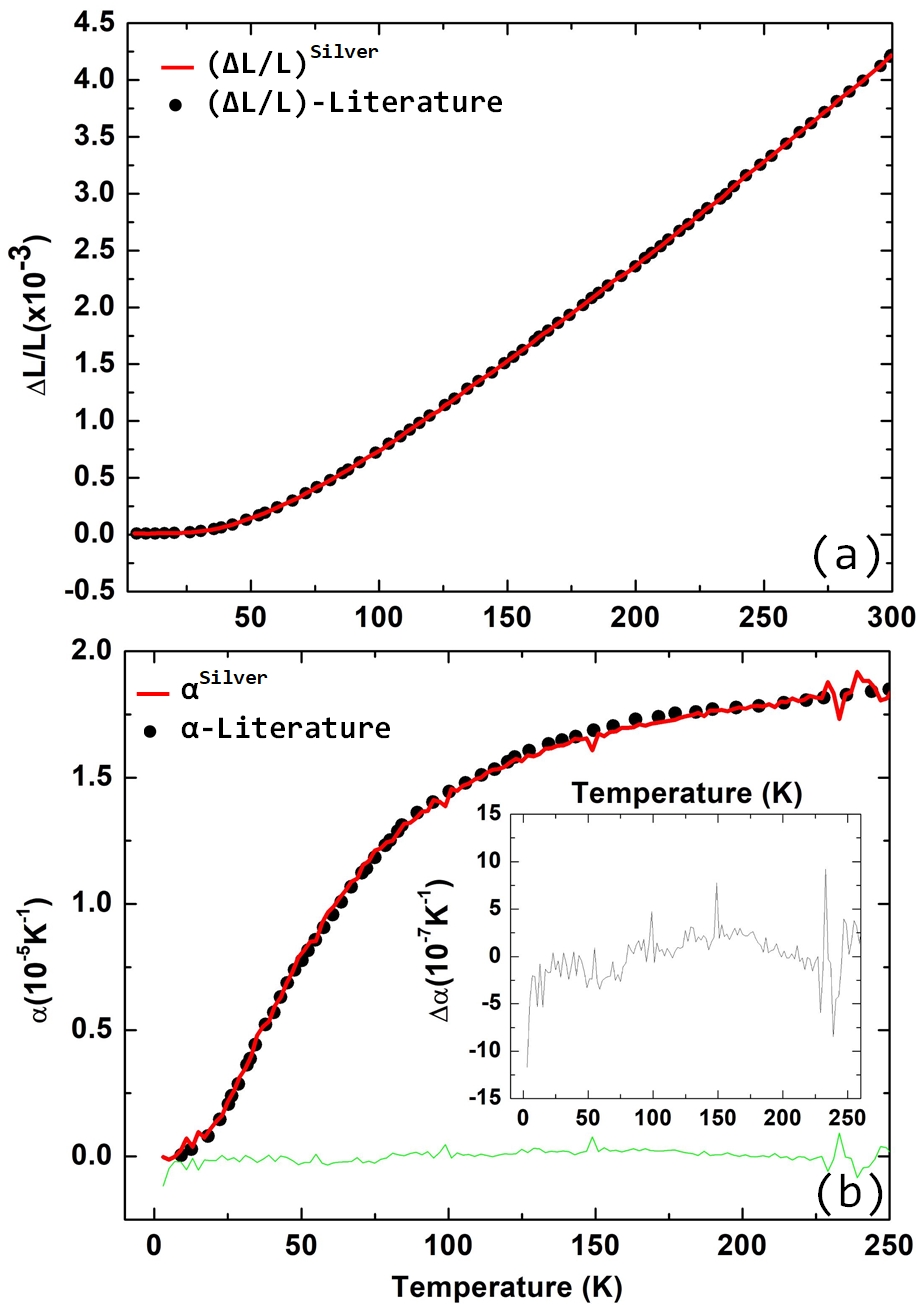}
		\caption{(Colour online) (a) Measured (red) and the literature (black) values of the relative length change $\Delta L/L$ for a silver sample obtained on $dilatometer \#2$ using $CCR\#2$. Literature values were obtained from K{\"u}chler et al. \cite{kuchler}. (b) Red continuous curve denotes coefficient of linear thermal expansion $\alpha$ obtained on the data of (a) using the polynomial fitting algorithm, while the black filled circles correspond to literature values extracted from K{\"u}chler et al. \cite{kuchler}.}
		\label{fig:Silver}
	\end{center}
\end{figure}
$dilatometer \#2$ on $CCR\#2$ for thermal expansion measurements, we measured another metal with well known thermal expansion characteristics, namely, silver. Red continuous curve in Fig. \ref{fig:Silver}(a) shows the plot of $\Delta L/L$ obtained on a polycrystal of silver using $dilatometer \#2$ and measured on $CCR\#2$, while the black filled circles correspond to literature values extracted from \cite{kuchler}. From the plot, it is clear that our obtained values of $\Delta L/L$ match the literature data very well.\\
Figure \ref{fig:Silver} (b) shows the calculated linear thermal expansion coefficient $\alpha$ from the corresponding $\Delta L/L$ values of Fig. \ref{fig:Silver} (a) obtained by the polynomial fitting algorithm described above. It can be seen that the experimentally obtained $\alpha$ matches the literature data very well in the temperature range below 200 K. Above 200 K, there are couple of spikes in $\alpha$, possibly arising from the CCR.

\subsection{COMSOL simulation of uniaxial force exerted on a sample by the dilatometers}
In both the dilatometers, $dilatometer \#1$ as well as $dilatometer \#2$, the movement of the movable capacitor plate towards or away from the upper capacitor plate is achieved via a spring motion (spring $k$ in $dilatometer \#1$ and spring $K$ in $dilatometer \#2$). This results in an inevitable force, and consequently, a pressure on the sample. It is possible that the application of such a force may result in a non-monotonicity in the thermal expansion coefficient data, which may affect the measurements artificially. So, in order to check if a monotonic variation of displacement is achieved on application of forces with varying magnitudes, we performed a finite element method (FEM) simulation on various digital models of the spring, having a geometry similar to that used in $dilatometer \#1$ \cite{schmiedeshoff}. The simulation was done using COMSOL Multiphysics simulation software package which is a software cross-platform for finite element analysis, equation solving and multi-physics simulation \cite{comsol}. To build the various models of the circular spring \cite{schmiedeshoff} needed as an input to the COMSOL software, we used CREO Parametric software. For a particular model, the displacement field at various points was obtained by (i) uploading the model with desired thickness and diameter, (ii) various parametric curves drawn to divide the surface into regions, (iii) specifying the boundary condition and boundary load, and (iv) choosing a mesh size depending on the precision of the measurement-finer the mesh, more accurate the results.

\subsubsection{$Dilatometer\#1$}
\begin{figure}[H]
	\begin{center}
		\includegraphics[width=1.0\textwidth]{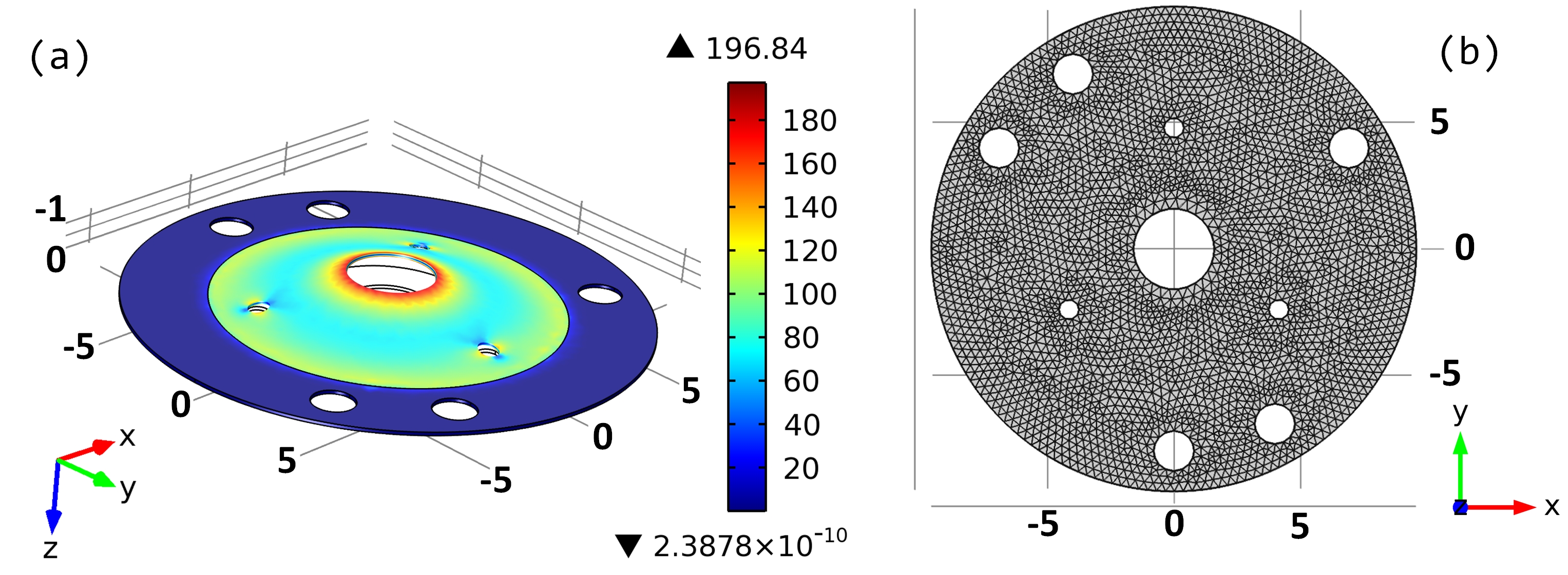}
		\caption{(Colour online) (a) FEM simulation done in COMSOL multiphysics software for an applied force of 0.2 N$/$m$^2$ at the centre of the spring by the sample. Colour code represents the variation of von Mises stress on the spring due to the application of the stress at the centre with blue representing the lowest value and dark red representing the highest. The resultant displacement of the spring is seen as a curvature such that the curvature is highest at the centre and lowest at the edge. (b) Representation of the mesh chosen for each element of the spring in order to calculate forces, displacement and spring constant of the spring.}
		\label{fig:Schematics_FEM}
	\end{center}
\end{figure}

\begin{figure}[H]
\begin{center}
	\includegraphics[width=1\textwidth]{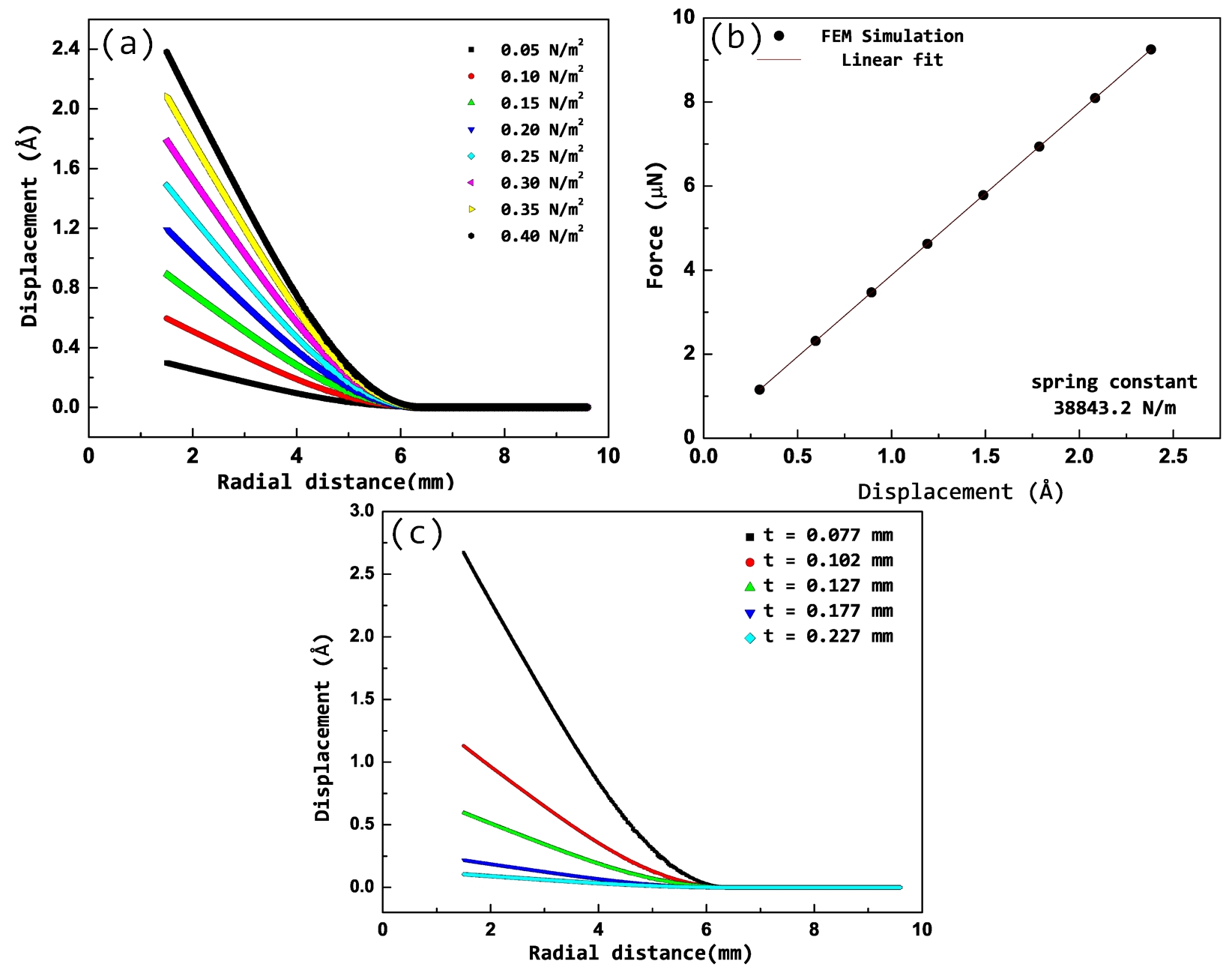}
	\caption{(colour online) (a) Variation of displacement of the spring of thickness 0.127 mm along the radial direction in the basal plane as a function of different values of pressures applied over an area of 23.127 mm$^2$. (b) Maximum displacement (co-ordinate of the center of the spring) vs. the applied force on the sample. Black dots represent the data from graph (a) and the red line represents a linear fit to the data. The slope of the graph gave the spring constant as 38843 N/m. (c) Deformation of the BeCu springs of thickness 0.077 mm (black), 0.102mm mm (red), 0.127 mm (green), 0.177 mm (blue), 0.227 mm (black) under same applied force of \textbf{1N}.}
	\label{fig:Force_Dependence}
\end{center}
\end{figure}

We started off by finding the displacement of the spring under the application of pressure (force per unit area), with values chosen in the range that produce displacements in the angstrom range. The application of a pressure results in von Mises stress in different parts of the spring \cite{vonmises}. It was found that application of large pressures of the order of 1000 N$/$m$^{2}$ resulted in the spring getting deformed. However,  in the low range of applied pressure of the order of 0.05 N$/$m$^{2}$ - 0.40 N$/$m$^{2}$, resulting in displacements of the order of few angstroms, the spring was well within the elastic limit, causing only elastic deformations. Fig. \ref{fig:Schematics_FEM} (a) shows a representative image of the deformation caused on the dilatometer spring in the z direction as well as the variation of the von Mises stress on the application of a pressure of magnitude 0.2 N$/$ m$^2$. The material input for the simulation was beryllium-copper (Be-Cu) alloy UNSC17200. This is the same alloy that we have used in fabricating our $dilatometer \#1$. Schmiedeshoff et al. \cite{schmiedeshoff} have also used the same material in their dilatometer. A single hole at the centre of the spring represents the hole made in the spring for holding the lower capacitor plate via the nut (refer to Fig. \ref{fig:Schematics_Instrumentation} (c)). The holes at the circumference of the circular spring correspond to the holes made in the spring to screw the spring to the cell frame $a$ by three $m$ screws, shown upside down in Fig. \ref{fig:Schematics_Instrumentation} (c). For the FEM simulation, this region is constrained to be fixed and is one of the boundary conditions that was applied, i.e. the applied pressure is zero at the position of the holes. The other boundary condition is that the maximum pressure is applied at the central region of the spring via a contact force that gets applied on the spring when the sample expands. Accordingly, the area of the holes at the circumference of the spring is shown as dark blue region and the central hole region is shown as dark red in Fig. \ref{fig:Schematics_FEM} (a). The region between these two colours (dark blue and dark red) shows the distribution of von Mises stress on application of a pressure in the central region.\\
The mesh size for the FEM calculation was dynamically allocated depending on the curvature of the surface, namely, a fine mesh was chosen near the holes and the edges, while a coarse mesh was chosen at other places. Each element of the FEM was chosen as a tetrahedron with a minimum element size of 0.686 mm and a maximum element size of 3.814 mm. The x and y co-ordinates in Figs. \ref{fig:Schematics_FEM} (a) and (b) denote the spring size in mm to represent the diameter of the Be-Cu spring which is $\sim$ 19 mm. The z co-ordinate in Fig. \ref{fig:Schematics_FEM} (a) is in the scale of 10$^{-7}$ mm, to reflect the displacement of the spring that occured in the spring on being subjected to forces via the sample.\\ 
Figure \ref{fig:Force_Dependence} (a) shows the displacement of the spring  that was subjected to pressures described in Fig. \ref{fig:Schematics_FEM} (a) above. The resultant displacement, plotted as a function of the radial distance, occurs as a result of the application of the pressure in the basal plane of the spring. It can be seen that the displacement is the maximum at 1.5 mm from the centre of the spring corresponding to the area where the pressure is applied. As one moves away from the centre, the displacement decreases and becomes $\sim$ zero at 6 mm. This is the position where the nuts are placed. From Fig. \ref{fig:Force_Dependence} (a), it is clear that the displacement varies linearly with radial distance in the range 1.5 mm to 5 mm for all values of applied pressures. Fig. \ref{fig:Force_Dependence} (b) shows the variation of the obtained displacement with respect to the applied force. It can be seen that the variation is clearly linear which is confirmed by a straight line fit, shown as red line in Fig. \ref{fig:Force_Dependence} (b). The slope of the fit gave the spring constant as 38843 N/m, in agreement with values obtained on springs made of similar material \cite{kuchler}.   

In order to see the effect of thickness of the spring on a resultant displacement, we simulated a displacement vs. radial distance curve for varying thicknesses of the spring, starting with 0.227 mm and ending with 0.077 mm. Fig. \ref{fig:Force_Dependence} (c) plots this curve where it is very clear that the minimum displacement obtained at a given radial distance is the minimum for the thickest spring of 0.227 mm while the maximum displacement is obtained for the thinnest spring of 0.077 mm. However, it is also to be noted that the displacements obtained by the 0.102 mm spring is also quite large. We operate our dilatometer with this thickness of 0.1 mm. A lower thickness spring can give higher displacements but is extremely difficult to machine and may also result in easy deformations, so we fixed our thickness to 0.1 mm.
 
\subsubsection{$Dilatometer\#2$}
\begin{figure}[H]
	\begin{center}
		\includegraphics[width=1\textwidth]{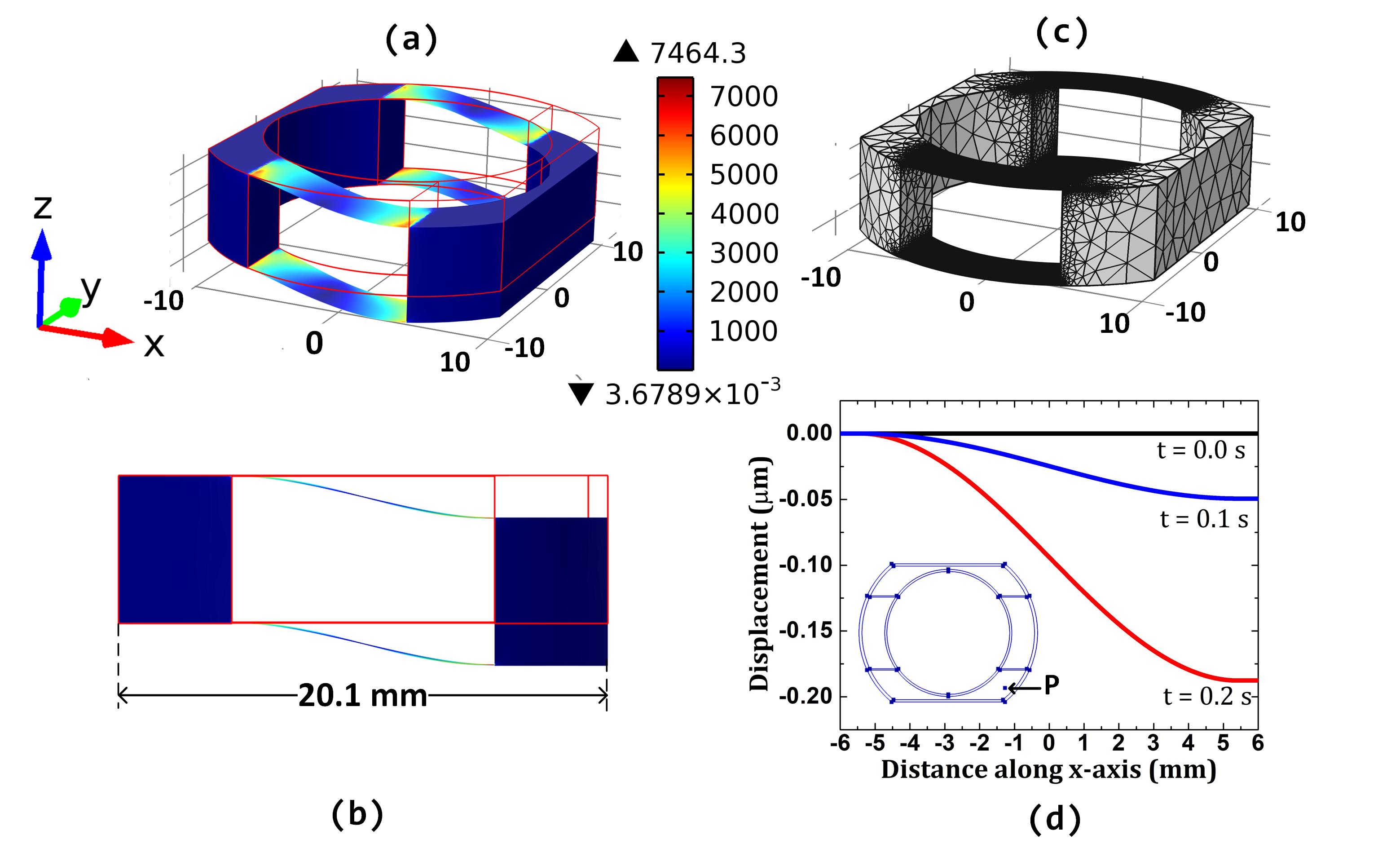}
		\caption{(Colour online) (a) Variation of von Mises stresses in the flange-spring unibody design due to the application of a force at point P shown in (d). Colour code represents the magnitude of the von Mises stress with blue denoting the minimum value and red denoting the maximum. (b) Representation of the meshes chosen for each element of the spring-unibody system for performing the FEM simulation. (c) Cross-sectional view of the movement of the leaf spring-flange unibody on the application of the force at point P. (d) Variation of displacement of the spring-flange unibody system with respect to the distance along the x-axis at time t = 0 s (black curve), 0.1 s (blue curve) and 0.2 s (red curve).}
		\label{fig:Force_Dependence_kuchler}
	\end{center}
\end{figure}

As described in the "Experimental setup" section above, $dilatometer\#2$ was machined out of a single block of BeCu and contained the main cell body, the two capacitor plates and the leaf springs. The mechanical constraint placed by such a design of the spring constrains the movable capacitor plate to move in only one direction offering very high parallelism between the two capacitor plates \cite{steinitz,kuchler,neumeier}. In order to demonstrate this, we have performed FEM simulations on a model of the $dilatometer\#2$. The input drawings for the simulations were made using CREO Parametric software. The material chosen for the simulation was the same Beryllium-copper (Be-Cu) alloy as specified by Schmiedeshoff et al. \cite{schmiedeshoff}, namely, UNSC17200. The first objective of the simulation was to show the parallel movement of the spring even under the application of a force applied at an asymmetric point P. So, the first boundary condition of the simulation is maximum force application at point P (shown in Fig. \ref{fig:Force_Dependence_kuchler} (d)). The second boundary condition is zero force at the left blue shaded area of Figs. \ref{fig:Force_Dependence_kuchler} (a) and (b). Parallel movement of the spring was found at the application of very small forces of the order of few $\mu$N as well as large values of forces. For calculation of the spring constant, the value of the force was chosen to be in conformity to the values of forces that were applied by K{\"u}chler et al. \cite{kuchler} in their actual measurement of the relative change in length brought about by the application of a force in $dilatometer\#2$. The resultant variation of von Mises stresses on the leaf spring due to the application of this pressure is shown in Fig. \ref{fig:Force_Dependence_kuchler} (a).\\
The mesh that were chosen for the entire flange-spring unibody system to execute the FEM simulation were dynamically allocated and are shown in Fig. \ref{fig:Force_Dependence_kuchler} (b). Each element of the mesh was chosen as a tetrahedron with the size of the largest element chosen at the cell body as 3.814 mm while at the leaf spring the finest mesh was chosen as 0.686 mm. The simulation was performed using a time dependent solver in the COMSOL software, where a fixed constraint was applied to the part opposite to point P. Fig. \ref{fig:Force_Dependence_kuchler} (c) shows the cross-sectional view of the resultant displacement of the leaf spring-flange unibody system. It can be clearly seen from the figure that the application of an asymmetric force at a corner of the cell body results in a parallel movement of the leaf spring-flange unibody without any bend. Fig. \ref{fig:Force_Dependence_kuchler} (d) shows the time evolution of the obtained displacement in $\mu$m as a function of the distance along the x-axis. It is to be noted that our obtained values of displacement are in the same order of magnitude ($\mu$m) as that obtained in \cite{kuchler}. The simulation was performed for a time of 0.2 s. As can be seen from Fig. \ref{fig:Force_Dependence_kuchler} (d), the displacement of the spring-flange unibody is maximum at the point of application of the force ($\sim$ 6 mm away from the centre of the x-axis) and goes to zero at $\sim$ -6 mm from the centre of the x-axis. The values of the maximum displacement obtained at the point P increases as a function of time, as expected.\\
\begin{figure}[H]
	\includegraphics[width=1\textwidth]{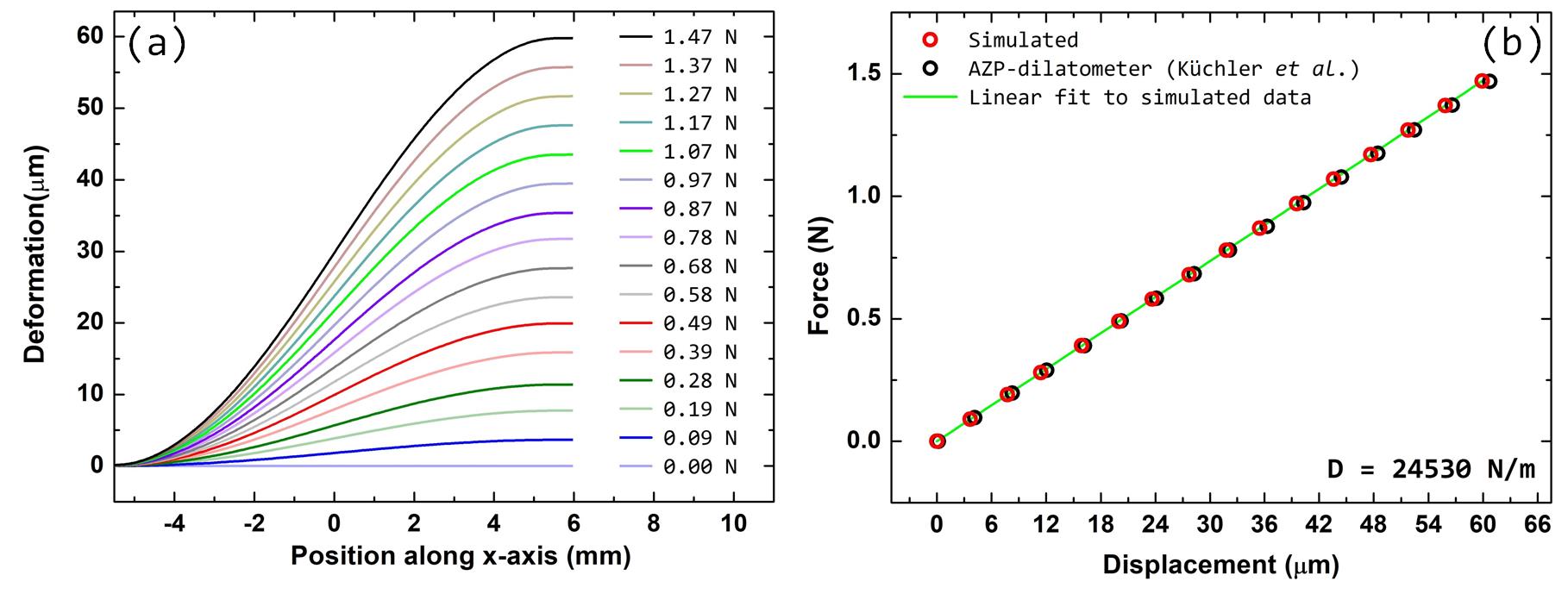}
	\caption{(colour online) (a) Displacement of Be-Cu leaf spring of thickness 0.25 mm as a function of applied force varying from 0.09 N to 1.47 N. The step change of applied force was 0.01 N. (b) Variation of applied force with the maximum displacement obtained in (a). Red open circles denote the data obtained from the simulation while the black open circles represent extracted data from \cite{kuchler}. Green solid line is a straight line fit to the simulated data. D = 24530 N$/$m represent the slope of the fit.}
	\label{fig:kuchler displacement}
\end{figure}

In order to see the effect of the variation of applied force on the displacement of the leaf spring, we simulate displacement vs. distance curves as shown in Fig. \ref{fig:kuchler displacement} (a). The values of applied forces ranged from 0.09 N (dark blue) to 1.47 N (black). From Fig. \ref{fig:kuchler displacement} (a), it can be clearly seen that the displacement rises from zero value at $\sim$ -6 mm away from the centre of the x-axis and reaches a maximum at +6 mm. The value of obtained displacement at a given position along the x-axis also increases with an increase in the value of applied force, in exact conformity with the results obtained for $dilatometer\#1$ (c.f. Fig. \ref{fig:Force_Dependence} (a)). Red open circles in Fig. \ref{fig:kuchler displacement} (b) represent the maximum value of displacement for a given value of applied force obtained from Fig. \ref{fig:kuchler displacement} (a). The values of applied forces were chosen to be of similar values as the one used by K{\"u}chler et al. in their force vs. displacement curve (c.f. Fig. 4 of \cite{kuchler}). Black open circles in Fig. \ref{fig:kuchler displacement} (b) denote the extracted values of displacement from reference \cite{kuchler}. It can be seen that our simulated values match those obtained from \cite{kuchler} very well and the variation of force with displacement is linear. A straight line fit to the data, shown as green solid line in Fig. \ref{fig:kuchler displacement} (b), gave the slope as 24530 which corresponds to the spring constant of the material. The obtained value of the spring constant is in excellent agreement with the value 24291 N$/$m obtained by K{\"u}chler et al. \cite{kuchler}.

\subsection{Thermal expansion measurements on  YBa$_2$Cu$_{3-x}$Al$_x$O$_{6+\delta}$ and Bi$_2$Sr$_2$CaCu$_2$O$_{8+x}$}
After describing the details of the two closed-cycle cryostats and the two dilatometer's functioning on standard metal samples, we measured thermal expansion on single crystals of two different high temperature superconductors, namely, YBa$_2$Cu$_{3-x}$Al$_x$O$_{6+\delta}$ (Al-YBCO) and Bi$_2$Sr$_2$CaCu$_2$O$_{8+x}$(BSCCO-2212). The measurements were done on $dilatometer\#2$ on $CCR\# 2$, i.e. on K{\"u}chler's dilatometer and dynacool closed cycle refrigerator. The single crystal of Al-YBCO selected for the measurements was an as-grown crystal that was grown by a self-flux method employing a vertical temperature gradient and had a superconducting onset temperature T$_c^{on}$ = 58.5 K \cite{manjuYBCO}. The as-grown crystal was not twinned and no detwinning was applied on the crystals before the thermal expansion measurements in order to avoid any defect incorporation in the crystals \cite{kund}. The length of the crystal along the c-axis is 1.02 mm.
\begin{figure}[H]
	\begin{center}
		\includegraphics[width=0.7\textwidth]{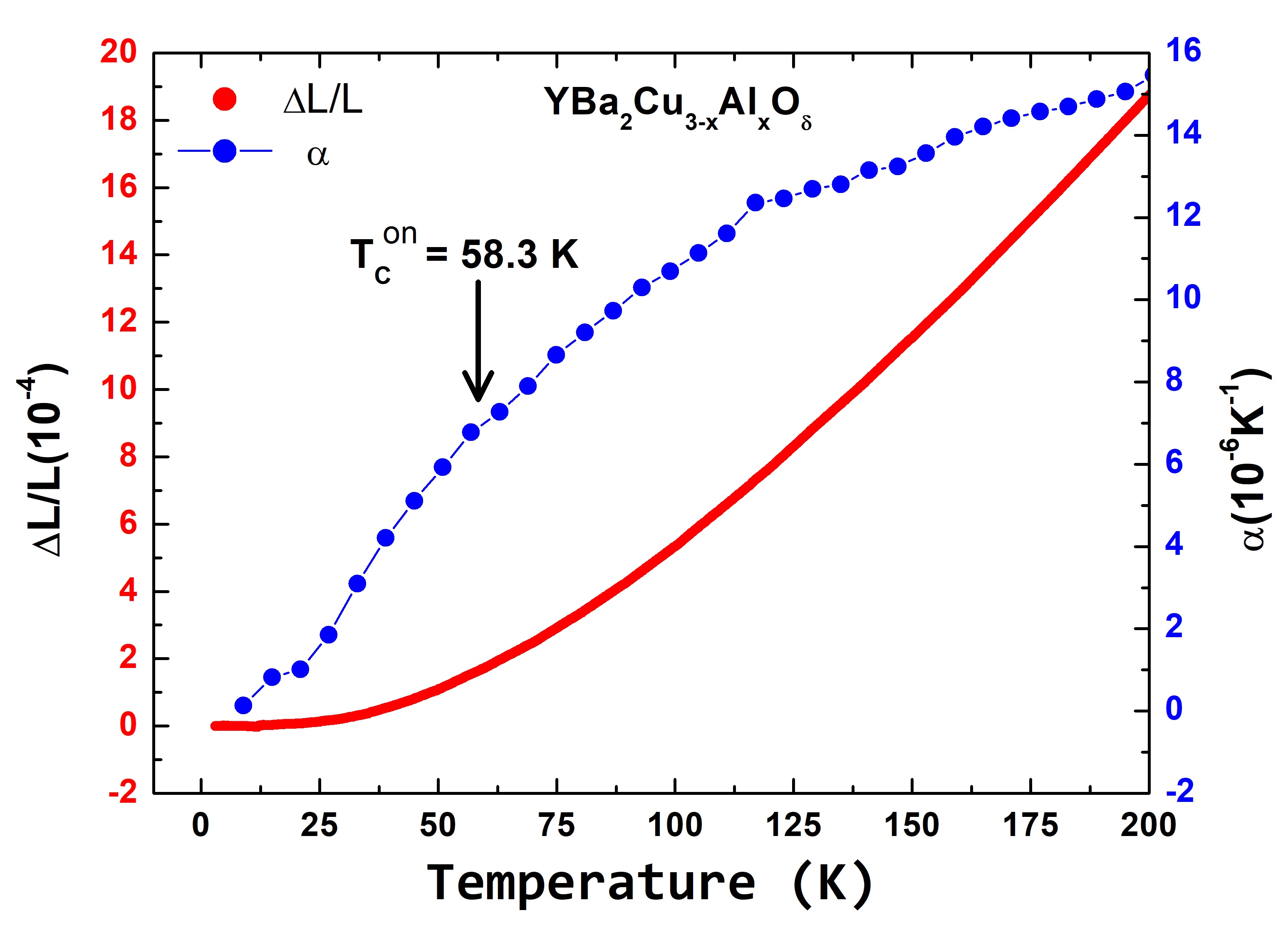}
		\caption{(colour online) Red filled circles denote $\Delta L/L$ values obtained on an as-grown single crystal of YBa$_2$Cu$_{3-x}$Al$_x$O$_{6+\delta}$ (Al-YBCO) measured along the c-axis. Red labels on the left represent the corresponding $\Delta L/L$ values. Blue filled circles denote the linear thermal expansion coefficient $\alpha$ obtained on the same Al-YBCO single crystal by using the polynomial fit on $\Delta L/L$ data. Blue labels on the right represent the corresponding $\alpha$ values. Onset of superconducting transition temperature T$_c^{on}$ is marked by a vertical arrow.}
		\label{fig:YBCO}
	\end{center}
\end{figure}
Red filled circles in Fig. \ref{fig:YBCO} denote $\Delta L/L$ values obtained on the single crystal of Al-YBCO measured along the c-axis while the blue filled circles denote the corresponding thermal expansion, $\alpha$. The data was obtained by following the polynomial fit method described above. So, a second order polynomial was chosen for the fit with the number of data points fitted simultaneously, N = 60. It can be seen that the $\alpha$ curve shows a smooth variation with temperature, in agreement to other reports of temperature variation of $\alpha$ measured along the c-axis in Al-YBCO \cite{kund, kund1,meingast,lortz}. It is very well known that the thermal expansion data measured along the c-direction of Al-YBCO does not give any jump at $T_c$ but only a smooth curve, as observed. Guided by the magnetisation data \cite{manjuYBCO}, we have marked the onset temperature of superconductivity, $T_c^{on}$, as 58.3 K.\\

Finally, we present thermal expansion data on the single crystal of another high temperature superconductor BSCCO-2212. The single crystal was grown by the self-flux regrowth method and had a T$_c^{on}$ of 88.3 K. The details of the crystal growth and characterisation of its structural and superconducting properties can be found in \cite{rajak,rajak1}. The measurements were done for the length along the crystallographic c-axis and the chosen crystal had a total length of 1.23 mm along c-axis.

\begin{figure}[H]
	\begin{center}
		\includegraphics[width=0.7\textwidth]{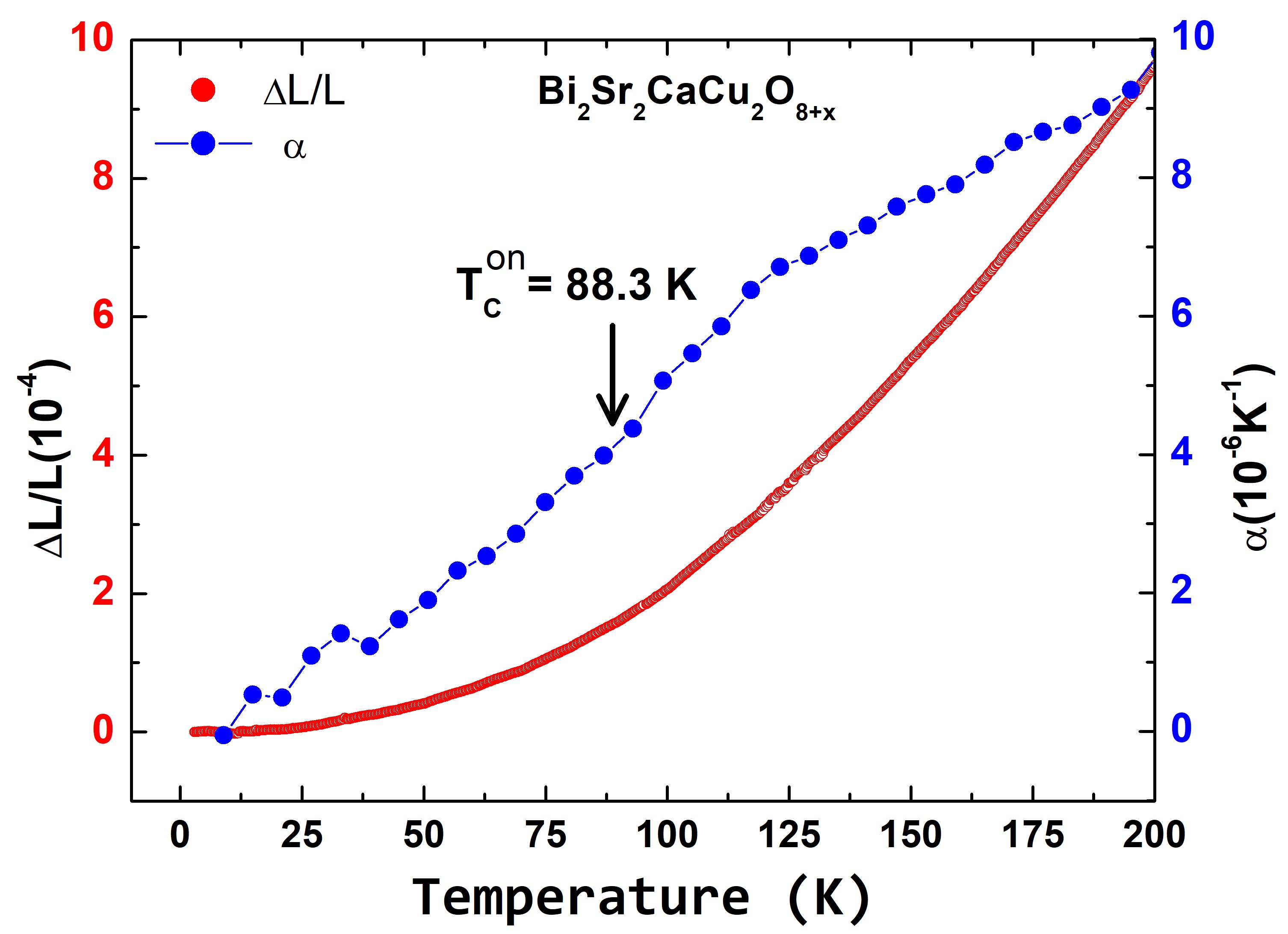}
		\caption{(colour online) Measured values of $\Delta L/L$ obtained on Bi$_2$Sr$_2$CaCu$_2$O$_{8+x}$ superconductor shown in red colour. Linear thermal expansion coefficient $\alpha$ obtained by using a polynomial fit on the corresponding $\Delta L/L$ data is shown in blue. The labels on the right are the corresponding $\alpha$ values. Onset of superconducting transition temperature T$_c^{on}$ is marked by a vertical arrow.}
		\label{fig:BSCCO}
	\end{center}
\end{figure}

Similar to the data on Al-YBCO, red filled circles in Fig. \ref{fig:BSCCO} denote the $\Delta L/L$ values for BSCCO-2212 crystal while the blue filled circles denote the linear thermal expansion coefficient $\alpha$ extracted from the corresponding $\Delta L/L$ values. The values were obtained with a polynomial fit of second order and N = 60. Our values of $\Delta L/L$ and $\alpha$ match quite well with other reported values \cite{meingastbscco,mukherjee}. Similar to the observation in Al-YBCO, the linear thermal expansion coefficient $\alpha$ is found to vary smoothly with temperature without any discontinuities \cite{meingastbscco,mukherjee}. We have marked the onset temperature of superconductivity T$_c^{on}$ as 88.3 K, found from magnetisation measurements \cite{rajak}.

\section*{Conclusion}
To conclude, we have demonstrated the utility of closed-cycle cryostats to perform thermal expansion measurements which are known to be extremely sensitive to noise vibrations. We employed two different kinds of dilatometers for the thermal expansion measurements, one that was built in-house and the other that was procured commercially. We tested the functionality of the two dilatometers on two commercially available closed cycle cryostats one of which works on the principle of Gifford-McMahon cooling while the second employs pulse-tube cooling technology for achieving the closed-cycle refrigeration. By a series of thermal expansion measurements performed on polycrystals of standard metals, aluminium and silver, we demonstrate excellent data, similar to those reported in wet cryostats. We show two different techniques of obtaining linear thermal expansion $\alpha$, namely, numerical integration and polynomial fit and find polynomial fits to give better data. The obtained $\alpha$ were found to be in excellent agreement to those obtained using wet liquid helium based cryostats. Between the two, pulse tube based cryostats give relatively better data than Gifford-McMahon based cryostat. We performed finite element method simulations on both the dilatometers to understand the spring action motion and the resultant movement of the capacitor plate. This was done to ensure that length change of the sample that produces an equivalent force/pressure on the spring makes the spring still work in the elastic limit. Using the simulations, spring constant of both the springs were found that had excellent match with published data. Finally, we measured thermal expansion on single crystals of two high temperature superconductors, YBa$_2$Cu$_{3-x}$Al$_x$O$_{6+\delta}$ and Bi$_2$Sr$_2$CaCu$_2$O$_{8+x}$ along the crystallographic c-axis and found very good match with published data obtained using wet cryostat, demonstrating great technological possibilites for future thermal expansion measurements using closed cycle cryostats.     

\section*{Acknowledgement:}
The authors acknowledge R. Suryanarayanan for fruitful scientific discussions. D. J-N acknowledges financial support from SERB-DST, Govt. of India (Grant No. YSS/2015/001743). 

\section*{Data availability}
The data that support the findings of this study are available from the corresponding author upon reasonable request.

\end{document}